\documentclass[usenatbib]{mn2e}
\usepackage{graphicx}
\usepackage{lscape}
\usepackage{color}

\begin{document}

\title[Galactic discs mass transport.]{Multiscale mass transport in 
z$\sim$6 galactic discs: fueling black holes.}

\author[Prieto \& Escala]{Joaquin Prieto$^{1}$\thanks{email:joaquin.prieto.brito@gmail.com}\& Andr\'{e}s 
Escala$^{1}$\\ 
$^1$ Departamento de Astronom\'{i}a, Universidad de Chile, Casilla 36-D, Santiago, Chile.}

\maketitle

\begin{abstract}
By using AMR cosmological hydrodynamic N-body zoom-in simulations, with 
the RAMSES code, we studied the mass transport processes onto galactic nuclei
from high redshift up to $z\sim6$. Due to the large dynamical range of 
the simulations we were able to study the mass accretion process on scales 
from $\sim50[kpc]$ to $\sim$ few $1[pc]$. We studied the BH growth on to the 
galactic center in relation with the mass transport processes associated 
to both the Reynolds stress and the gravitational stress on the disc. 
Such methodology allowed us to identify the main mass transport process as 
a function of the scales of the problem. We found that in simulations that 
include radiative cooling and SNe feedback, the SMBH grows at the 
Eddington limit for some periods of time presenting 
$\langle f_{EDD}\rangle\approx 0.5$ throughout its evolution. The 
$\alpha$ parameter is dominated by the Reynolds term, $\alpha_R$, with 
$\alpha_R\gg 1$. The gravitational part of the $\alpha$ parameter, 
$\alpha_G$, has an increasing trend toward the galactic center at higher redshifts, 
with values $\alpha_G\sim 1$ at radii $\la$ few $ 10^1[pc]$ contributing to the BH fueling. 
In terms of torques, we also found that gravity has an increasing 
contribution toward the galactic center at earlier epochs with a mixed contribution
above $\sim 100 [pc]$. This complementary work between 
pressure gradients and gravitational potential gradients allows an 
efficient mass transport on the disc with average mass accretion rates 
of the order $\sim$ few $1 [M_{\odot}/yr]$. These level of SMBH accretion rates 
found in our cosmological simulations are needed in all models of SMBH growth 
that attempt to explain the formation of redshift $6-7$ quasars.
\end{abstract}


\section{Introduction}
The mass transport (MT) process in astrophysical environments has relevance for different
phenomena in our Universe. It is important for planet formation in proto-planetary
discs, it triggers the AGN activity associated to super massive black hole (SMBH) 
accretion at high redshift and it is responsible for mass accretion from the 
filamentary structures around dark matter (DM) haloes into the central regions of the 
first galaxies. A full understanding of this phenomenon is very important 
in the construction of a galaxy formation theory.

In particular, the MT phenomenon has a crucial relevance in models of black hole (BH) 
formation and their growth in the early stages of our Universe. The observation of very 
bright quasars at redshift $z\ga6$ with luminosities $L\ga10^{13}[L_\odot]$ implies the 
existence of BHs with masses of the order $M_{BH}\sim10^{9}[M_\odot]$ when our Universe 
was about $\sim 1$ Gyr old \citep{Fan+2001}, i.e. SMBH should be formed very early in 
the history of our Universe and they should grow very fast in order to reach such high 
masses in the first $\sim Gyr$ of our Universe. To understand such a rapid early evolution 
is one of the main challenges of current galaxy formation theories. For a more extended 
discussion on massive BH formation at high redshift see \citet{Volonteri2010} and 
\citet{Haiman2013}.

There are three main scenarios for the formation of SMBH seeds:
\begin{itemize}
\item Seeds from the first generation of stars and their subsequent accretion to form
the SMBH: In this scenario a top heavy initial mass function (IMF) associated with 
population III (pop III) stars \citep{Abel+2002,BrommLarson2004} can leave BHs with masses of 
the order of $\sim100$M$_\odot$ \citep{HegerWoosley2002}. We note that more recent
studies have shown that pop III stars could be formed in clusters of low mass stars, e.g.
\citet{Stacy+2010,Greif+2011,Clark+2011}.
\item Massive seeds formed by the direct collapse of warm ($\ga 10^4$ K) neutral hydrogen 
inside atomic cooling haloes: In this scenario the initial BH seeds are formed by the direct 
collapse of warm gas triggered by dynamical instabilities inside DM haloes of mass 
$\ga5\times 10^7[M_\odot]$ at high redshift, $z\ga10$ \citep[e.g. ][]{OhHaiman2002,LodatoNatarajan2006,Begelman+2006}. 
If there is no molecular hydrogen
to avoid fragmentation \citep[e.g. ][]{Agarwal+2012,Latif+2013,Latif+2014} and there is an efficient 
outward AM transport \citep{Choi+2015} such a scenario could favor the formation of a BH seed of 
$\sim10^4-10^6[M_\odot]$, e.g. \citet{Begelman+2008}.
\item BH seeds formed by dynamical effects of dense stellar systems like star clusters or
galactic nuclei, e.g. \citet{Schneider+2006}: In this scenario the formed BH seed 
can be of mass $\sim10^2-10^4[M_\odot]$ \citep{DevecchiVolonteri2009}
\end{itemize}

Studies related to the initial mass for SMBH formation favor massive 
$\sim 10^4-10^6$M$_\odot$ seeds inside primordial atomic cooling haloes
\citep{Volonteri+2008,TanakaHaiman2009,LodatoNatarajan2006}. Despite that,
due to the lack of observational evidence it is not clear yet if one of these
scenarios is preferred by nature or all of them are working at the same time
in different haloes. 

There is dynamical evidence for the existence of SMBH in the center of nearby galaxies 
\citep{Ferrarese&Ford2005} with masses in the range $M_{BH}\sim10^6-10^9[M_\odot]$ 
suggesting that the BHs formed in the first evolutionary stages of our Universe 
are now living in the galactic centers around us, including our galaxy \citep{Ghez+2005}. 
Besides their ubiquitous nature there is evidence of scaling relations connecting the
BH mass with its host galaxy properties, namely the galactic bulge - BH mass relation
\citep[e.g. ][]{Gultekin+2009} and the bulge stars velocity dispersion - BH mass relation
\citep[e.g. ][]{Tremaine+2002,FerrareseMerritt2000}. 
Such relations suggest a co-evolution between the BH and its host galaxy. 

In a cosmological context, motivated by the theoretical study of 
\citet{PichonBernardeau1999} a series of recent 
simulations \citet{Pichon+2011} and \citet{Codis+2012} have argued that the 
galactic spin may be generated entirely in the baryonic component due to the growth 
of eddies in the turbulence field generated by large-scale ($\ga$ few Mpc) mass 
in-fall relating the large scale angular momentum (AM) acquisition with mass 
transport phenomena inside the virial radius. 
 
\citet{Danovich2015} studied the AM acquisition process in galaxies 
at redshift $z\approx 4-1.5$ identifying 4 phases for AM acquisition: i) the initial spin 
is acquired following the Tidal Torque Theory 
\citep{Peebles1969,Doro70}. ii) In this phase both the mass and AM are transported to the 
halo outer region in the virialization process following the filamentary structure
around them. iii) In the disc vicinity the gas forms a non-uniform ring which
partially suffers the effect of the galactic disc torques producing an alignment 
with the inner disc AM. iv) Finally, outflows reduce gas with low specific AM, increasing 
its global value at the central region and violent dynamical instabilities (VDI)
associated to clump-clump interactions and clump-merged DM halo interactions remove AM, 
allowing a more centrally concentrated gas. 

At smaller scales ($\sim$ few $100 [kpc]$), \citet{PrietoSpin} studied the MT and AM 
acquisition process in four DM haloes of similar mass $M\approx10^9[M_\odot]$ and 
very different spin parameter $\lambda=0.001-0.04-0.06-0.1$ \citep{Bullock2001} 
at redshift $z=9$. The main result of this work is the anti-correlation between
the DM halo spin parameter and the number of filaments converging on it: the larger
the number of filaments the lower the spin parameter. Such a result suggests that DM haloes
associated to isolated knots of the cosmic web could favor the formation of SMBH
because the inflowing material would have to cross a lower centrifugal barrier to 
reach the central galactic region.

In a non-cosmological context, \citet{Escala2006} and \citet{Escala2007} has shown 
that the interplay and competition between BH feeding and SF can naturally explain 
the $M_{BH}-\sigma$ relation. Using idealized isolated galaxy evolution simulations 
\citet{Bournaud+2007} showed that $z\sim 1$ galaxies are able to form 
massive clumps due to gravitational instabilities \citep{Toomre1964} triggered by its 
high gas mass fraction. Such clumpy high redshift galaxies evolve due to 
VDI to form spiral galaxies characterized by a bulge and an exponential disc. Similar 
results have been found in cosmological contexts by \citet{Mandelker+2014}. They 
show that the formation of massive clumps is a common feature of $z\sim 3-1$ galaxies.
Due to the fast formation and interaction between them the VDI dominate the disc evolution. 
A similar clump migration has been observed at higher redshift in a $5\times10^{11}[M_\odot]$ 
DM halo at $z=6$ in \citet{Dubois+2012} and \citet{Dubois+2013}. In these works the migration 
has been triggered by DM merger induced torques. In contrast to the scenario presented 
above there are studies supporting the idea that clump interaction in high redshift
discs are not the main source to build up the galactic bulges 
\citep[e.g. ][]{Hopkins+2012,Fiacconi+2015,Tamburello+2015,Behrendt+2016,Oklopcic+2016}.  

Inspired by the $\alpha$ parametrization in the seminal paper 
of \citet{SS1973}, \citet{Gammie2001} studied the gravitational stability 
in cool thin discs. In his work \citet{Gammie2001} quantified the rate of angular momentum 
flux in terms of the Reynolds and gravitational stress. In this work we will 
use a similar $\alpha$-formalism to study the MT process on galactic discs 
at redshift $z\sim 6$ performing N-body and hydrodynamic numerical simulations from cosmological 
initial conditions. We will study a halo of $\sim$ few $10^{10}[M_\odot]$, a mass value not
studied already in this context. It is the first time that such an approach is being
used to study the MT process on galaxies at high redshift. Furthermore we will 
compute directly the torques working on the simulated structures from $\sim50 [kpc]$
scales associated to the cosmic web around the central DM halo to $\sim$few pc scales
associated to the galactic disc. Such an approach will allow us to have clues about 
the main source of mass transport on these objects and then to have some insights about
the SMBH growth mechanisms at high redshift.

The paper is organized as follows. Section \S 2 contains the numerical details of our 
simulations. Here we describe the halo selection procedure, our refinement strategy and 
the gas physics included in our calculations. In section \S 3 we show our results. Here 
we present radial profiles of our systems, star formation properties of our galaxies, a 
gravitational stability analysis and show a mass transport analysis based on both the 
$\alpha$ formalism and the torques analysis on small and large scales. In section 
\S 4 we discuss our results and present our main conclusions. 

\section{Methodology and Numerical Simulation Details}
\label{Methodology}

\subsection{RAMSES code}
The simulations presented in this work were performed with the cosmological N-body
hydrodynamical code RAMSES \citep{Teyssier2002}. This code has been written to study
hydrodynamic and cosmological structure formation with high spatial resolution using 
the Adaptive Mesh Refinement (AMR) technique, with a tree-based data structure. The 
code solves the Euler equations with a gravitational term in an expanding universe 
using the second-order Godunov method (Piecewise Linear Method). 

\subsection{Cosmological parameters}
Cosmological initial conditions were generated with the mpgrafic code 
\citep{Prunetetal2008} inside a $L=10[cMpc]$ side box. Cosmological parameters where
taken from \citet{Planck2013Results}: $\Omega_m=0.3175$, $\Omega_\Lambda=0.6825$, 
$\Omega_b=0.04899$, $h=0.6711$, $\sigma_8 =0.83$ and $n_s=0.9624$.

\subsection{Halo selection}
Using the parameters mentioned above, we ran a number of DM-only simulations with 
$N_p=256^3$ particles starting at $z_{ini}=100$. We selected one DM halo of mass 
$M_{DM}\approx 3\times 10^{10}[M_\odot]$ at redshift z=6. We gave preference to DM 
haloes without major mergers through its final evolution in order to have a more 
clean and not perturbed system to analyze.

After the selection process we re-simulated the halo including gas physics. For 
these simulations we re-centered the box on the DM halo position at redshift 
$z=6$. We set a coarse level of $128^3$ (level 7) particles and allowed for further 
DM particle mass refinements until level 10 inside a variable volume called mask 
(as we wil explain below). In this way we were able to reach a DM 
resolution equivalent to a $1024^3$ particle grid inside 
the central region of the box, which corresponds to a particle mass resolution 
$m_{part}\approx3\times10^4[M_\odot]$, in other words we resolved the high redshift 
$\sim10^6[M_\odot]$ halo with $\ga30$ particles and our final halo with $\ga 10^6$ particles. 

\subsection{Refinement strategy}
In order to resolve all the interesting regions we allowed refinements
inside the Lagrangian volume associated to a sphere of radius $R_{ref}=3R_{vir}$\footnote{Here 
$R_{vir}\equiv R_{200}$, the radius associated to an spherical over-density 200 
higher than $\Omega_m\rho_c$, with $\rho_c$ the critical density at the corresponding 
redshift.} around the selected DM halo at $z_{end}=6$. Such a Lagrangian volume is tracked 
back in time until the initial redshift of the simulation, $z_{ini}$=100. In this way we
ensure that the simulation is resolving all the interesting volume of matter throughout 
the experiment, i.e. all the material ending inside the $R_{ref}$ at the end of 
the simulation\footnote{In order to define such a volume we compute a mask. Such a mask 
can be computed using the geticref.f90 and the geticmask.f90 routines in the 
${\rm /ramses/utils/f90/zoom\_ic}$ folder.}. The Lagrangian volume (the mask) is defined 
by an additional passive scalar advected by the flow throughout the simulation. At the 
beginning the passive scalar has a value equal to 1 inside the mask and it is 0 outside. 
We apply our refinement criteria in regions where this passive scalar is larger than $10^{-3}$.  

In our simulations a cell is refined if it is in regions where the mask passive scalar is
larger than $10^{-3}$ and if one of the following conditions is fulfilled:
\begin{itemize} 
\item it contains more than 8 DM particles, 
\item its baryonic content is 8 times higher than the average in the whole box, 
\item the local Jeans length is resolved by less than 4 cells \citep{Trueloveetal1997}, and 
\item if the relative pressure variation between cells is larger than 2 (suitable for 
shocks associated to the virialization process and SNe explosions). 
\end{itemize}

Following those criteria the maximum level of refinement was set at 
$\ell_{max}=18$, corresponding to a co-moving maximum spatial resolution of 
$\Delta x_{min}\approx38.1 [cpc]$ and a proper spatial resolution of 
$\Delta x_{min}\approx5.4 [pc]$ at redshift $z=6$. With this resolution we were
able to resolve the inner $0.1R_{vir}$ DM region with $\sim200$ computational
cells.

\subsection{Gas physics}
Our simulations include optically thin gas cooling. The gas is able to cool due to H, 
He and metals following the \citet{SutherlandDopita93} model until it reaches a temperture of
$T=10^4 [K]$. Below this temperature the gas can cool until $T=10[K]$ due to metal lines cooling.
We note that the cooling functions assume collisional ionization equilibrium. 
The metals are modeled as passive scalars advected by the gas flow. In order to mimic the 
effect of H$_2$ cooling in primordial environments all our simulations started with an initial
metallicity $Z_{ini}=0.001[Z_\odot]$ \citep{Powelletal2011}. Furthermore, a uniform 
UV background is activated at $z_{reion}=8.5$, following \citet{HaardtMadau1996}. 

\subsubsection{Star formation}
The numerical experiments include a density threshold Schmidt law for star formation:
above a given number density, set to $n_0\approx 30[cm^{-3}]$ in our case, the gas is 
converted into stars at a rate density, $\dot{\rho}_\star$, given by \citep[e.g. ][]{RaseraTeyssier2006,DuboisTeyssier2008}:
\begin{equation}
\dot{\rho}_\star=\epsilon_\star \frac{\rho}{t_{ff}(\rho)}
\end{equation}
where $\rho$ is the local gas density, $\epsilon_\star=0.05$ is the constant star
formation efficiency and $t_{ff}(\rho)$ is the density dependent local free fall time 
of the gas. The number density for star formation, $n_0$, corresponds to the value at 
which the local Jeans length is resolved by 4 cells with a temperature $T_0=200 [K]$\footnote{In 
order to avoid numerical fragmentation we added a pressure floor to the hydrodynamical pressure. 
The pressure floor is computed as
\begin{equation}
P_{floor}=\frac{\rho k_{B}}{m_H}\frac{T_{floor}}{\mu},
\end{equation}
with
\begin{equation}
\frac{T_{floor}}{\mu}=T_0\left(\frac{n}{n_0}\right).
\end{equation}
The pressure floor is activated at $n_0$ for $T_0$ and at the corresponding density for different
temperatures. We note that under the Jeans condition $T_{floor}\propto n$ and $P_{floor}\propto n^2$.}

When a cell reaches the conditions to form stars we create star particles following a
Poisson distribution
\begin{equation}
P(N)=\frac{\lambda_P^N}{N!}e^{-\lambda_P},
\end{equation}
with $N$ the number of formed stars and 
\begin{equation}
\lambda_P=\frac{\rho\Delta x^3}{m_\star}\frac{\Delta t}{t_\star},
\end{equation}
where $\Delta x$ is the cell grid side, 
$m_\star\approx m_H n_0 \Delta x^3$ is the mass of the stars, $\Delta t$ is
the time step integration and $t_\star=t_{ff}(\rho)/\epsilon_\star$ is the 
star formation time scale. This process ends up with a population of stars inside
the corresponding cell. In order to ensure numerical stability we do not allow 
conversion of more than $50\%$ of the gas into stars inside a cell.

\subsubsection{SNe feedback}
After 10 Myr the most massive stars explode as SNe. In this process a mass fraction 
$\eta_{SN}=0.1$ (consistent with a Salpeter initial mass function truncated between
$0.1$ and $100[M_\odot]$) of the stellar populations is converted into SNe ejecta:
\begin{equation}
m_{eject}=\eta_{SN}\times m_\star.
\end{equation}
In this case $m_\star$ is not the single stellar particle mass but the total stellar
mass created inside a cell. Furthermore, each SNe explosion a releases specific energy
$E_{SN}=10^{51}[erg]/10 [M_\odot]$ into the gas inside a sphere of $r_{SN}=2\Delta x$:
\begin{equation}
E_{eject}=\eta_{SN}\times m_\star \times E_{SN}.
\end{equation}

As mentioned above, metals are included as passive scalars after each SNe explosion
and then they are advected by the gas flows. This means that after each SNe explosion 
a metallicity
\begin{equation}
Z_{eject}=0.1[Z_\odot]
\end{equation}
is included as metals in the gas in the simulation. Such an amount of metals is consistent 
with the yield of a $10[M_\odot]$ type II SNe from \citet{WoosleyWeaver95}.

In this work we used the delayed cooling implementation of the SNe feedback 
\citep[discussed in][]{Teyssier+2013,Dubois+2015}. This means that in places where 
SNe explode, if the gas internal energy is above an energy threshold $e_{NT}$, 
the gas cooling is turned off for a time $t_{diss}$ in order to take 
into account the unresolved chaotic turbulent energy source of the explosions.

As written in \citet{Dubois+2015} the non-thermal energy $e_{NT}$ evolution 
associated to the SNe explosions can be expressed as
\begin{equation}
\frac{d e_{NT}}{dt}=\eta_{SN}\dot{\rho}_\star E_{SN}-\frac{e_{NT}}{t_{diss}}=\eta_{SN}\epsilon_\star\rho \frac{E_{SN}}{t_{ff}}-\frac{e_{NT}}{t_{diss}}.
\end{equation}
In an equilibrium state $de_{NT}/dt=0$ it is possible to write
\begin{equation}
\frac{e_{SN}}{\rho}=\eta_{SN}\epsilon_\star E_{SN}\frac{t_{diss}}{t_{ff}}.
\end{equation}
If we assume that non thermal energy is associated to a turbulent motion with a
velocity dispersion $\sigma_{NT}$ and that this energy $e_{NT}=\rho\sigma_{NT}^2/2$ 
will be dissipated in a time scale of order the crossing time scale associated to 
the local jeans length then $t_{diss}\approx l_J/\sigma_{NT}$, and it is possible to 
write
\begin{equation}
t_{diss}=\left(\frac{t_{ff}}{2\eta_{SN}E_{SN}\epsilon_\star}\right)^{1/3}l_J^{2/3}.
\end{equation}
Then, expressing the local Jeans length as $l_J=4\Delta x$, with $\Delta x$
the proper cell side at the highest level of refinement, it is possible to write the 
dissipation time scale as:
\begin{eqnarray}
t_{diss} & \approx & 0.52[Myr]\left(\frac{0.1}{\eta_{SN}}\right)^{1/3}\left(\frac{0.05}{\epsilon_\star}\right)^{1/3}\times \\ \nonumber
         &   & \left(\frac{\Delta x}{5.4[pc]}\right)^{2/3}\left(\frac{30[cm^{-3}]}{n_0}\right)^{1/6}
\end{eqnarray}
Given our parameters, we set the non-thermal energy dissipation time scale 
as $t_{diss}\approx0.5[Myr]$.

In this model the gas cooling is switched off when the non-thermal velocity dispersion
is higher than a given threshold:
\begin{eqnarray}
\sigma_{NT}& \approx & 49[km/s]\left(\frac{\eta_{SN}}{0.1}\right)^{1/3}\left(\frac{\epsilon_\star}{0.05}\right)^{1/3}\times \\ \nonumber
           &   & \left(\frac{\Delta x}{5.4[pc]}\right)^{1/3}\left(\frac{n_0}{30[cm^{-3}]}\right)^{1/6}
\end{eqnarray}
which for us is $\sigma_{NT} \approx 49 [km/s]$.

\subsection{Sink particles and black hole accretion}
In order to follow the evolution of a black hole (BH) in the simulations we introduced 
a sink particle \citep{Bleuler&Teyssier2014} at the gas density peak inside a DM halo 
of $M\approx1.7\times10^{8}[M_\odot]$ at redshift $z=15.7$. The BH seed mass is 
10$^4M_\odot$, roughly following the $M_{BH}-\sigma$ relation of \citet{McConell+2011}.
Such a black hole mass is in the range of masses associated to direct collapse 
of warm gas inside atomic cooling haloes at high redshift
\citep[e.g. ][]{OhHaiman2002,LodatoNatarajan2006,Begelman+2006,Begelman+2008,Agarwal+2012,Latif+2013,Latif+2014,Choi+2015}. 
We did not allow more BH formation after the formation of the first one. In order to 
compute the mass accretion rate onto the BH we use the modified Bondi-Hoyle accretion rate 
described below.

\subsubsection{Modified Bondi accretion} 
\citet{Bleuler&Teyssier2014} implemented the modified Bondi mass accretion rate onto
sink particles in the RAMSES code. They use the expression presented in 
\citet{Krumholz+2004} based on the Bondi, Hoyle and Lyttleton theory 
\citep{Bondi&Lyttleton1939,Bondi1952}. There the Bondi radius
\begin{equation}
  {r_{BHL}}=\frac{G{M}_{BH}}{(c^2_\infty+v^2_\infty)}
\end{equation}
defines the sphere of influence of the central massive object of mass $M_{BH}$ and 
its corresponding accretion rate is given by
\begin{equation}
	\dot{M}_{BHL}=4\pi\rho_\infty {r}^2_{BHL}(\lambda^2c^2_\infty+v^2_\infty)^{1/2}
\end{equation}
where $G$ is the gravitational constant, $c_\infty$ is the average sound speed and 
$v_\infty$ is the average gas velocity relative to the sink velocity, $\lambda$ is 
an equation of state dependent variable and it is ${\rm exp}(3/2)/4\approx1.12$ in 
the isothermal case. The density $\rho_\infty$ is the gas density far from the central
mass and it is given by
\begin{equation}
\rho_\infty=\bar{\rho}/\alpha_{BHL}(\bar{r}/r_{BHL})
\end{equation}
where $\alpha_{\rm BHL}(x)$ is the solution for the density profile in the Bondi model \citep{Bondi1952}.
The variable $x=\bar{r}/r_{\rm BHL}$ is the dimensionless radius and $\bar{\rho}$ the
corresponding density. In our case $\bar{r}=2\Delta x_{min}$, with $\Delta x_{min}$ the
minimum cell size in the simulation. 

The modified Bondi accretion rate is limited by the Eddington accretion rate. In 
other words, the sink particle can not accrete at a rate larger than the Eddington 
rate, given by:
\begin{equation}
	\dot{M}_{Edd}=\frac{4\pi G m_p M_{BH}}{\sigma_T c\epsilon_r},
\end{equation}
where $m_p$ is the proton mass, $\sigma_{\rm T}$ is the Thomson scattering cross
section, $c$ is the speed of light and $\epsilon_r=0.1$ is the fraction of accreted 
mass converted into energy. 

\section{Results}
\label{results}

In this work we will analyze three simulations: 
\begin{itemize}
\item NoSNe simulation: Includes star formation and modified Bondi-Hoyle-Lyttleton 
(BHL) accretion rate onto sinks, 
\item SNe0.5 simulation: Includes star formation, BHL accretion rate onto sinks and 
SNe feedback with a consistent delayed cooling $t_{diss}=0.5[Myr]$, and
\item SNe5.0 simulation: Similar to SNe0.5 but with an out of model $t_{diss}=5.0[Myr]$ 
\end{itemize}
We are not including AGN feedback in these experiments. This important
ingredient has been left for a future study to be presented in an upcoming publication.

Figure \ref{fig:mapsim} shows a gas number density projection of our systems at redshift
$z=6$ at two different scales. The top rows show the large scale 
($\sim 3\times10^2 [ckpc]$) view of the systems and the bottom rows show a zoom-in of the 
central ($\sim 10 [ckpc]$) region.

In the top panels it is possible to recognize a filamentary structure converging on 
to the central galaxy position. Such filaments work as pipes channeling cold baryonic 
matter onto the converging region: the place for galaxy formation, i.e. 
knots of the cosmic web. Aside from the accreted low density gas it is possible to 
recognize a number of over-densities associated to small DM haloes merging 
with the central dominant halo, a common feature of the hierarchical structure
formation. Such mini haloes certainly perturb the galactic disc environment 
as we will see later. 

Whereas at large scale we can see that the difference between runs are the 
small scale features associated to the shocks produced by SNe explosions, the bottom 
panels show a clear difference between simulations at the end of the experiments. 
The NoSNe run developed a concentrated gas rich spiral galaxy whereas both SNe runs have 
less concentrated gas and much more chaotic matter distribution. Such a
difference certainly is a consequence of SNe explosions: the energy injected into 
the environment is able to spread the gas out of the central region and then to decrease 
the average density due to effect of the expanding SNe bubbles. Such a phenomenon is able
to destroy the galactic disc as can be seen in the central and right-bottom panel where 
the proto-galaxy is reduced to a number of filaments and gas clumps. 

\begin{figure*}
\centering
\includegraphics[width=2.0\columnwidth,height=11.5cm]{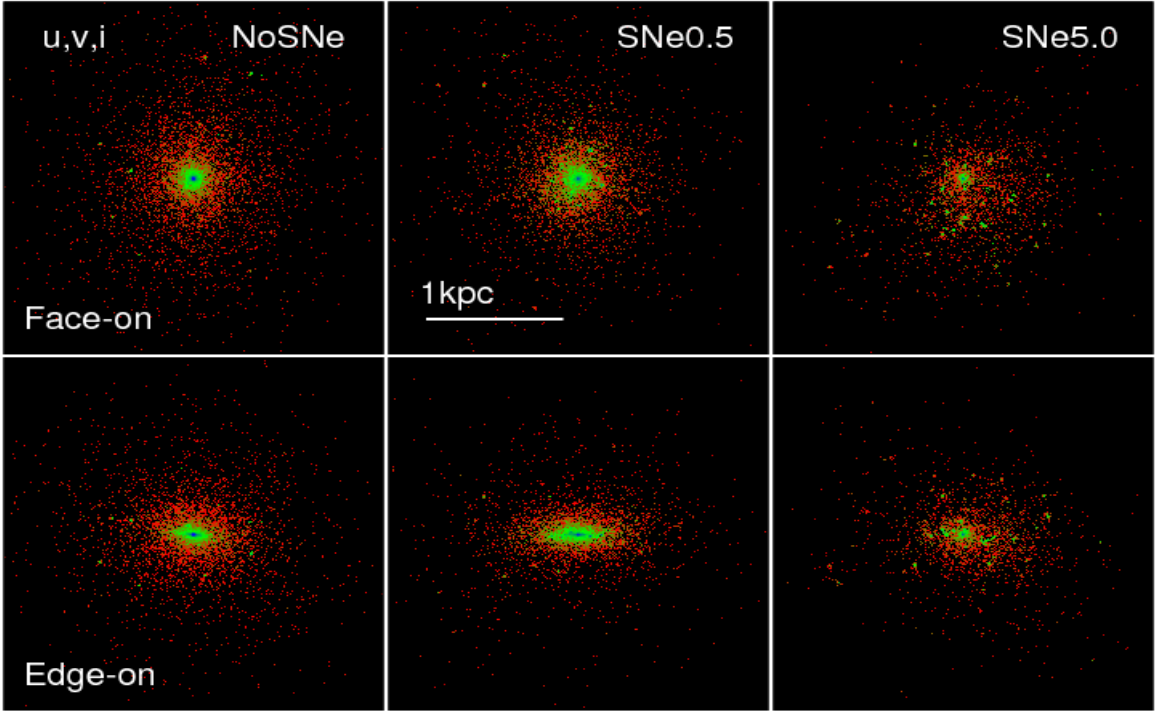}
\caption{Combined rest frame stars visualization for our three runs using SDSS $u$, $v$ and $i$ 
filters in blue green and red colors, respectively. The images correspond to the end of our 
simulations and there is no dust extinction. The face on view of the NoSNe system shows a 
smoother star distribution compared with our
SNe runs where feedback is able to create a non-homogeneous star distribution characterized by
green-blue star clumps around the center of the system.}
\label{fig:starmaps}
\end{figure*}

Figure \ref{fig:starmaps} shows the rest frame face on (top panels) and edge-on (bottom panels)
stellar populations associated to our systems in three combined filters: i, u and v. The images 
were made using a simplified version of the STARDUST code \citep{Devriendt+99}\footnote{The 
STARDUST code computes the observed flux for a single stellar population (ssp). It assumes 
a Salpeter IMF and for a given stellar track it computes the associated spectral 
energy distribution of a ssp. Then it is convolved with the different filters from 
SSDS to obtain the observed maps. There is no dust extinction in our case.}.

\subsection{Radial profiles}

Before computing any radial average from our AMR 3D data we aligned the 
gas spin vector with the Cartesian $\hat{z}$ direction. After this procedure, 
we performed a mass weighted average in the $\hat{z}$ direction in order to have 
all the interesting physical quantities associated to the disc surface:
\begin{equation}
\langle Q(x,y)\rangle_z=\frac{\sum_{z_i}^{z_f} Q(x,y,z) \Delta m}{\sum_{z_i}^{z_f}\Delta m},
\end{equation}
with $\Delta m=\rho \Delta x^3$ and $\Delta x$ the grid size.

In order to get our radial quantities we have performed a mass weighted averaging 
in the cylindrical $\hat{\theta}$ direction of our disc surface data:
\begin{equation}
\langle Q(r)\rangle_\theta=\frac{\sum_{\theta=0}^{\theta=2\pi} Q(r,\theta) \Delta m}{\sum_{\theta=0}^{\theta=2\pi}\Delta m},
\end{equation}
where $r$ and $\theta$ are the radial and azimuthal cylindrical coordinates, with 
$\Delta m=r \Delta r \Delta\theta$ and $\Delta r=\Delta x$.

\begin{figure}
\centering
\includegraphics[width=1.0\columnwidth]{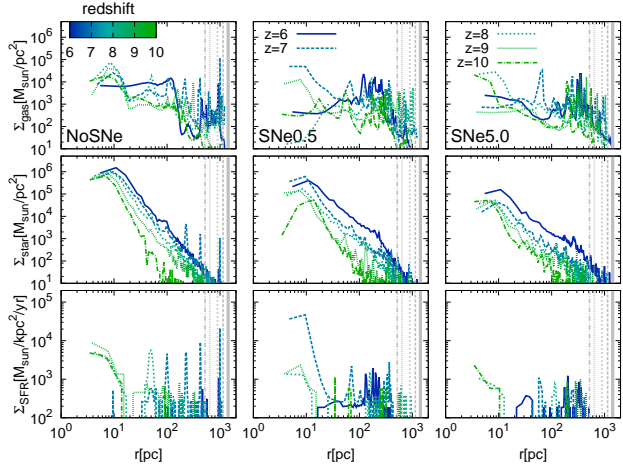}
\caption{Radial profiles of the gas SD (top panels), stars SD (central panel) and SD SFR 
(bottom panel) for the NoSne run in the left column, SNe0.5 run in the central column
and SNe5.0 run in the right column. All the quantities are plotted for different redshifts
(from 6 in blue to 10 in green): $z=6$ (solid line), $z=7$ (long-dashed line), 
$z=8$ (short-dashed line), $z=9$ (dotted line) and $z=10$ (dot-dashed line). The 
vertical lines mark the $4=0.1R_{vir}$ at each $z$ following the same line style as the 
profiles. The top and central panels show density fluctuations associated to the 
non-homogeneous nature of 
the galactic disc at high redshift. SNe feedback clearly decreases the amount of stars
formed in our galaxies.}
\label{fig:surfprof}
\end{figure}

Figure \ref{fig:surfprof} shows the gas surface density\footnote{In this case we have added
the total mass in the $\hat{z}$ direction (not an average) divided by the corresponding area
and then we averaged in the cylindrical direction in order to get mass weighted radial SD, for
gas, stars and SFR.} (SD) (top panels),
the stars SD (central panel) and the SD star formation rate (bottom panel) for different 
redshifts as a function of radius. (The line style-redshift relation will be kept for the
following plots). All our simulations show a number of peaks in the
gas SD profiles at almost all the sampled redshifts, a proof of the irregular and
clumpy structure of the gas.

The second row of the figure shows more clear differences between our runs. Apart from 
an almost monotonic increment on the stars SD for each simulation it is also
possible to see lower central peaks from left to right. Such a trend is a consequence 
of the different strength of the feedback which also increases from left to right.
 
The bottom panels show the SD star formation rate (SFR). In order to compute this 
quantity we took into account all the stars with an age $t_\star< 10 Myr$ formed in 
the time elapsed between consecutive outputs (which is in the range $\Delta t\sim7-9Myr$). 
Inside one tenth of the virial radius we compute the height and the radius where
$90\%$ of the gas and stars are enclosed. We averaged these scales for stars and gas
to define a cylinder where we compute the SFR (a similar procedure will be used
in the next sub-section to compute the Kennicutt-Schmidt law). We can see that our NoSNe run
shows higher SD SFR peaks above $\sim 10 pc$ and the SNe5.0 run has the lower SD SFR.
Such a fact will be confirmed after computing the global SD against the global SD SFR
in the next section.

\begin{figure}
\centering
\includegraphics[width=1.0\columnwidth,]{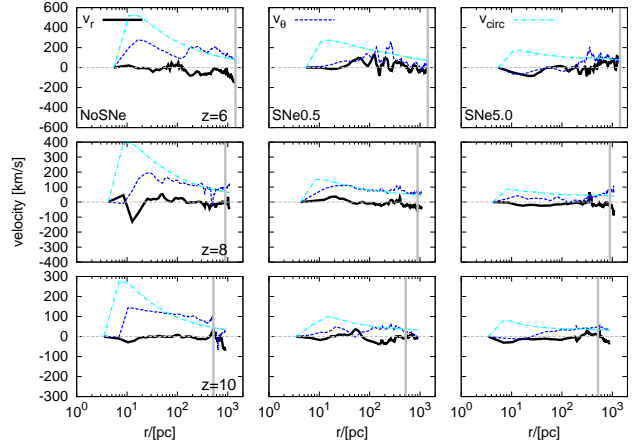}
\caption{The figure shows the gas radial velocity (solid black line), orbital velocity 
(long-dashed blue line) and the spherical circular velocity (dot-dashed cyan line)
as a function of radius at different redshifts for the NoSNe run (left columns), the
SNe0.5 run (central column) and the SNe5.0 run (right column). The vertical lines mark 
the $r=0.1R_{vir}$ position.}
\label{fig:velprof}
\end{figure}

Figure \ref{fig:velprof} shows different gas velocities associated to our systems at
different redshift as a function of radius. It plots the radial velocity 
$v_r\equiv\vec{v}\cdot \hat{r}$ (in solid black line), azimuthal velocity 
$v_\theta\equiv\vec{v}\cdot \hat{\theta}$ (in long-dashed blue line) and the 
spherical circular
velocity $v_{circ}$ of the disc (in dot-dashed cyan line) 
defined as:
\begin{equation}
v_{circ}=\left(\frac{GM(<r)}{r}\right)^{1/2}
\end{equation}
where $G$ is Newton's constant and $M(<r)$ is the total (gas, stars and DM) 
mass inside the radius $r$. 

\begin{figure*}
\centering
\includegraphics[width=2.0\columnwidth,height=1.0\columnwidth]{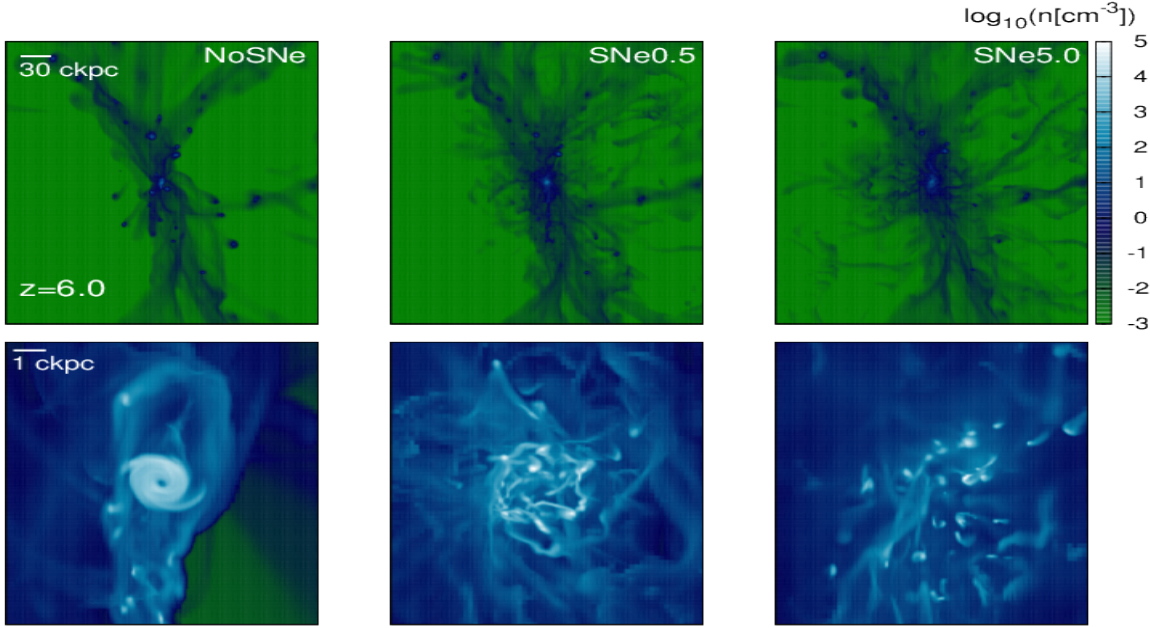}
\caption{Mass weighted projection of the gas number density for our simulations: 
NoSNe left column, SNe0.5 central column and SNe5.0 right column. The top row is 
a large scale ($\sim30$ ckpc square side) view of our systems and the bottom row is a zoom-in 
of the central region of the system ($\sim 1$ ckpc square side). From the top panels 
it is possible to identify the filamentary structure converging at the central region of 
the system: the galaxy position. Such filaments channel and feed the galactic structure. 
At large scales it is possible to recognize shock waves associated to the SNe explosions 
of our SNe runs. Beside the low density gas, there are a number of over-densities associated 
to small DM haloes about to merge with the central structure. The bottom panels show a 
dramatic difference between our simulations: a compact gas rich spiral galaxy for the
NoSne experiment, a rough spiral galaxy disturbed by SNe feedback in our SNe0.5 run and
a group of clumps in our SNe5.0 simulation.}
\label{fig:mapsim}
\end{figure*}

In our simulations the radial velocity fluctuates from negative to positive values.
Such a feature is a proof of a non stationary disc where at some radii there is 
inflowing material whereas at other radii there are gas outflows. Such features
can be produced by virialization shocks, DM halo mergers or SNe explosions. It is worth 
noticing that at $r\ga 1$ kpc, which roughly correspond to the outer edge of the galactic 
disc, the gas is inflowing in most of the cases. Such a feature is a remarkable signal of 
radial gas inflows at distance $\sim 0.1 R_{vir}$ from the center of the system. This fast 
radial material comes from larger scales channeled by the filamentary structure 
shown in the top panels of figure \ref{fig:mapsim} and, as mentioned above, they 
supply the central DM halo region with cold gas at rates as high as 
$\sim10[M_\odot/yr]$ as we will see in the following sections.

The orbital velocity tends to be roughly similar to the spherical circular velocity at large radii 
$r\ga 100[pc]$ in most of the cases but in general the circular velocity does not
follow the spherical circular orbit. Such deviations can be explained due to the 
shocked gas inflows, the mergers suffered by the central halo and due to SNe explosions which
enhance the pressure support against gravity. We emphasize that these kinds of interactions have 
a gravitational effect due to tidal forces (mergers and clump-clump interaction) on the disc 
and also have a hydrodynamical effect (shocks). In our SNe runs it is clear that the 
spherical circular curves are lower than the NoSNe curve. In other words the enclosed 
mass inside $\sim 0.1 R_{vir}$ is lower in the SNe runs. That is because the SNe explosions 
spread the gas out of the central region. Actually from the shocked gas features at the top 
right panel of figure \ref{fig:mapsim} it is possible to see that the outflows 
can reach regions at $\sim100 [ckpc]$ from the central galaxy, i.e. $\sim15 [kpc]$ at $z=6$.

\subsection{Star formation}
Despite the lack of observations of the Kennicutt-Schmidt (KS) law at high 
redshift ($z\ga 6$) it is interesting to compare the KS law from our simulations 
with its currently accepted functional form from \citet{Kennicutt98} (hereafter K98). 
Furthermore, it is also interesting to compare our data with more recent 
literature from the \citet{Daddi+10} (hereafter D10) results for normal and star 
burst galaxies.

\begin{figure}
\centering
\includegraphics[width=1.0\columnwidth]{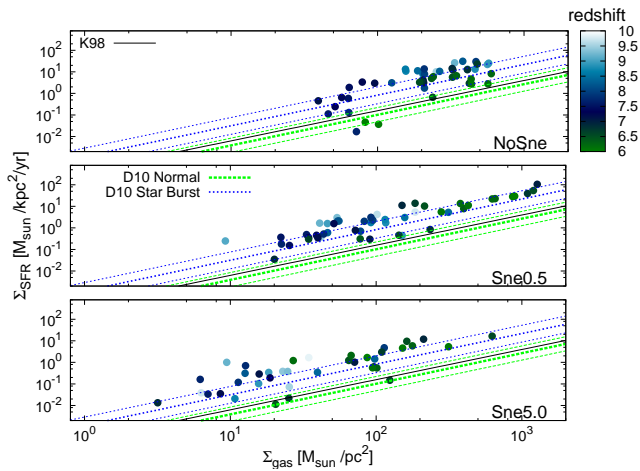}
\caption{KS law for each of our snapshots from $z=10$ to $z=6$. Each point marks the 
KS relation for our galaxies computed as mentioned in the text at different redshifts in
different colors. From top to bottom: NoSNe, SNe0.5 and SNe5.0. The solid black line 
marks the K98 fit. The thick long-dashed green line marks the D10 fit for normal galaxies 
and the short-dashed blue line marks the D10 fit for star burst galaxies. Our SNe0.5 run 
has the lowest scatter, and most closely follows the D10 sequence of star burst galaxies.}
\label{fig:ksl}
\end{figure}

Figure \ref{fig:ksl} shows the KS law for our runs. Each point marks the SD 
SFR as a function of the total gas SD. The SD SFR was computed following a similar procedure
as in the previous sub-section.

There is a correlation between the lower gas SD and the level of feedback in our results:
the higher the feedback the lower the gas SD, which is a natural consequence of the gas
heating due to SNe events. Whereas the NoSNe run shows points covering $\sim 1$ decade at
high SD with a large scatter in SD SFR from below the D10 normal galaxies sequence to above the 
D10 star burst sequence the SNe runs cover a larger SD range. Both SNe runs are in agreement with
the star burst sequence of D10 and the SNe0.5 simulation shows a lower scatter in the 
points. Such a behavior could be due to the number of mergers suffered by these kind of
haloes at high redshift.

\begin{figure}
\centering
\includegraphics[width=1.0\columnwidth]{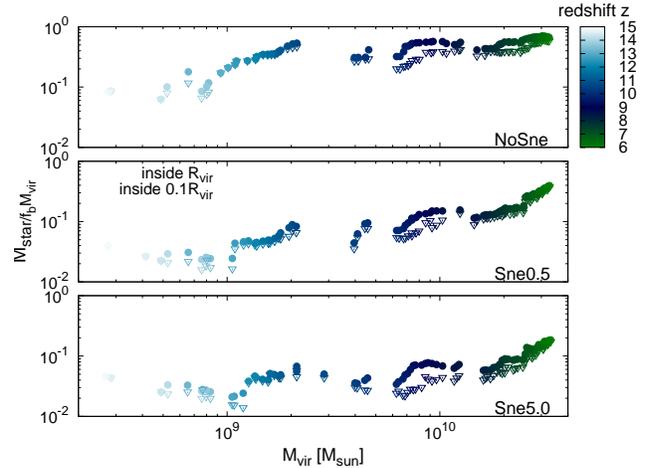}
\caption{Halo mass - stellar mass (normalized by the halo baryonic content assuming that 
it has exactly the universal baryonic fraction $f_b$) relation as a function of 
redshift in different colors for our runs. From top to bottom: NoSNe, SNe0.5 and SNe5.0. 
The filled circles mark the mass in stars inside the virial radius and the empty triangles 
mark the mass in stars inside one tenth of the virial radius. Our SNe5.0
run is in closest agreement with B13 at $\sim10^{10}M_\odot$ while for high $z$ galaxies our
SNe0.5 experiment is still in the limit of the order of magnitude predicted by B13. }
\label{fig:mvirmstar}
\end{figure}

Figure \ref{fig:mvirmstar} shows the stellar mass normalized by $f_b M_{vir}$, 
where $f_b\equiv \Omega_b/\Omega_m$ is the universal baryonic fraction. At the end
of the simulation our SNe0.5 galaxy has a stellar metallicity $Z_\star=0.1[Z_\odot]$
and our SNe5.0 galaxy a metallicity of $Z_\star=0.04[Z_\odot]$. It is clear from the figure that the 
NoSNe run is producing much more stars than our SNe runs and that due to the extreme
feedback our SNe5.0 simulation form less stars than our SNe0.5 simulation.
When we compare our results with the one shown in \citet{Kimm+2015} we can 
see that our results are in the range of their MFB and MFBm simulations at similar
$\sim10^{10}[M_\odot]$. Despite the uncertainties and the lack of robust observational 
constrains, such values are not far (a factor of $\sim$ few for SNe5.0
and just in the limit of the order of magnitude for SNe0.5) from the prediction 
from \citet{Behroozi+13} (hereafter B13) where the stellar to halo mass ratio is of the 
order of few $\sim 10^{-2}$ at the same mass range and high redshift.

\begin{figure}
\centering
\includegraphics[width=1.0\columnwidth]{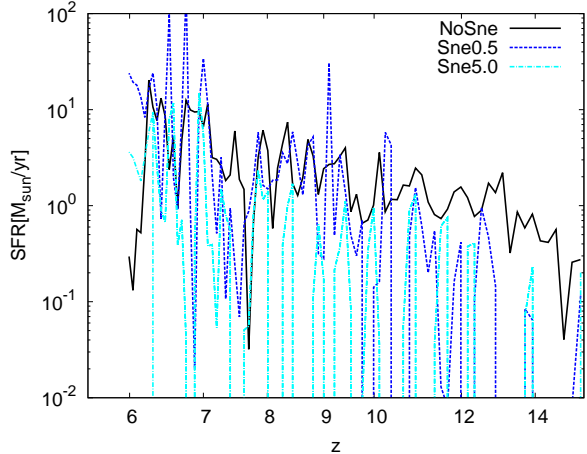}
\caption{SFR as a function of redshift for our experiments: NoSNe (black solid line),
SNe0.5 (dashed blue line) and SNe5.0 (dot-dashed line). It is worth to notice that above 
redshift $z\approx10$ the NoSne simulation presents a higher SFR compared with the SNe 
simulations: the NoSne experiment can form stars continuously without feedback. The SNe 
runs shows a ``bursty'' nature with peaks of SF in each $\sim$ few $10[Myrs]$.}
\label{fig:SFR}
\end{figure}

Figure \ref{fig:SFR} shows the SFR for our runs as a function of redshift. 
When we compare the NoSNe run with our SNe runs the main difference 
arises in the continuity of the SFR history. Whereas the NoSNe run shows a
continuous line the SNe runs present periods of almost zero SFR. Such periods 
last few $\sim 10[Myr]$ and are more frequent in our SNe5.0 run due to the
stronger feedback. Despite the large fluctuations in the SFR data our SNe 
runs tend to be in the range $\sim1-10[M_\odot/yr]$. Such numbers are in line
with the one found by \citet{Watson+15} (here after W15) and references therein
for high redshift galaxies. Taking into account the uncertainties of the predictions, 
if we compare our SNe run results with B13 they tend to be below or similar to their 
data for $\sim 10^{11}[M_\odot]$ halo at $z\ga6$.

\subsection{Disc stability}
\begin{figure}
\centering
\includegraphics[width=1.0\columnwidth]{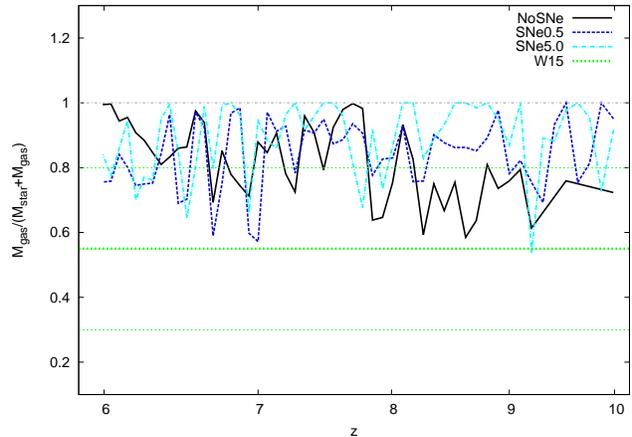}
\caption{Gas mass fraction as a function of redshift for our three runs: NoSNe (solid
black line), SNe0.5 (dashed blue line) and SNe5.0 (dot-dashed cyan line). The dotted 
thick green line at $f_g=0.55$ marks the W15 observed value. The dotted thin green line
are the errors associated to that observation. In general our SNe runs are well inside
the error bars of W15. Our SNe0.5 run has $f_g$ closer to the observed value at $z\la 7$
and it is just in the limit of $f_g=0.8$ when we average below $z=8$.}
\label{fig:gasfraction}
\end{figure}

High redshift galactic environments have a high gas fraction $f_g\ga 0.5$ 
\citep[e.g. ][]{Mannucci+2009,Tacconi+2010}. 
Figure \ref{fig:gasfraction} shows the gas fraction
of our systems as a function of redshift. Here we define the gas fraction as the 
ratio between the galactic gas mass and the mass of the gas plus the stars in the galaxy: 
$f_g\equiv M_{gas}/(M_{gas}+M_{star})$. All our systems shows a high gas fraction with
values fluctuating around $\sim 0.8$. In fact the average values for our simulations are
$\langle f_{g}\rangle_{NoSNe}=0.86$,
$\langle f_{g}\rangle_{SNe0.5}=0.83$ and $\langle f_{g}\rangle_{SNe5.0}=0.82$ below $z=8$.
If we average below redshift $7$ we find $\langle f_{g}\rangle_{NoSNe}=0.87$,
$\langle f_{g}\rangle_{SNe0.5}=0.80$ and $\langle f_{g}\rangle_{SNe5.0}=0.85$. 
It is interesting to compare such numbers with those found by W15. In this work the 
authors describe the properties of a $z\approx 7.5$ galaxy. The galaxy at this redshift 
has a gas fraction $f_g=0.55\pm0.25$, in other words our values of $f_g$ are inside the 
errors associated to their observations as we can see in figure \ref{fig:gasfraction} 
with the SNe runs closer to the observational expectations.  

The non-stationary and highly dynamic nature of the high gas fraction systems makes 
them susceptible to gravitational instabilities. In order to analyze the disc stability 
throughout its evolution we will use the Toomre parameter, $Q_T$, stability criterion 
\citep{Toomre1964}:
\begin{equation}
Q_T=\frac{c_s\Omega}{\pi G \Sigma}
\end{equation}

A convenient modification of the Toomre parameter to take into account the turbulent
velocity dispersion of the fluid has the form $Q_T=v_{rms}\Omega/\pi G \Sigma$. 
Despite the {\it ad hoc} modification of the parameter it is not straightforward to interpret 
the turbulent velocity dispersion of the gas as a source of pressure counteracting the gravity 
\citep{ElmegreenScalo2004}. This comes from the fact that this pressure term could only 
be defined in the case where the dominant turbulent scale is much smaller than the region 
under consideration, which is in fact not the case of the ISM. Rigorous analysis indeed 
shows that turbulence can be represented as a pressure only if the turbulence is produced
at scales smaller than the Jeans length \citep[micro-turbulence in ][]{Bonazzola+1992}.
Therefore the gravitational instability analysis is not strictly applicable with a 
turbulent pressure term that could stabilize and dampen all the substructure below
the unstable scale associated to $v_{rms}$.

\begin{figure}
\centering
\includegraphics[width=1.0\columnwidth]{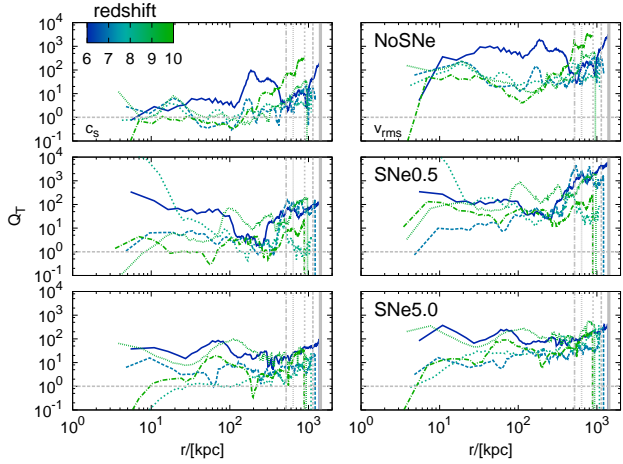}
\caption{Toomre parameter as a function of radius for our three simulations. The thermal
Toomre parameter in the left column and the turbulent Toomre parameter in the right column.
From top to bottom: NoSNe, SNe0.5 and SNe5.0. From the thermal Toomre panels we see that
our NoSNe run presents lower values compared with our SNe experiments. Despite that
the SNe runs do have some regions of the Toomre parameter $\la 1$ due to the effect of metal 
line cooling. Our turbulent Toomre parameter shows much higher values. This is due to the
high Mach number of our runs. The difference is more dramatic in our NoSNe run due to the
low gas temperatures reached without SNe heating. See the text for a discussion about this
parameter.}
\label{fig:toomreprof}
\end{figure}

The left column of figure \ref{fig:toomreprof} shows the Toomre parameter for our 
three runs at different redshifts. The gray dashed horizontal line marks the $Q_T=1$ state.
For completeness, the right column of figure \ref{fig:toomreprof} shows the Toomre 
parameter associated to the turbulent velocity dispersion. Due to the high Mach numbers 
(see appendix \ref{appendixD}) of these systems it is $\ga$ 1 order of magnitude above the 
thermal Toomre parameter.

Our NoSNe run tends to have lower values with a smaller dispersion compared with
our SNe runs. In the case of no feedback the Toomre parameter fluctuates around 1 above
$z=7$ showing an unstable disc at high redshift. At $z=6$ the parameter is of order $\sim10^0$
inside $\sim 100 pc$ and above this radius it increases due to the combined effect of
low density and higher sound speed (high temperature) stabilizing the system at these radii.
    
Due to the higher temperature associated to SNe explosions the Toomre parameter tends to
be larger in our SNe runs showing a more stable system in these cases. Despite that
it is also possible to find regions with $Q_T\approx 1$ in our feedback runs. We have
to take into account that after each SNe explosion a given amount of metals is released
into the gas. Such a new component allows the gas to reach lower temperatures creating
unstable regions. 

\begin{figure}
\centering
\includegraphics[width=1.0\columnwidth]{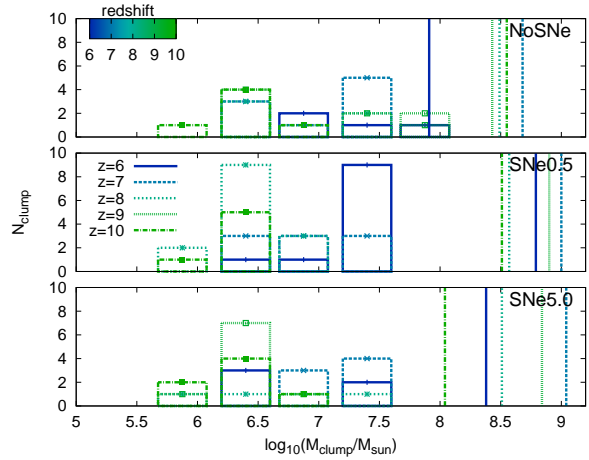}
\caption{The gas clumps mass function. From top to bottom: NoSNe, SNe0.5 and SNe5.0.
The vertical lines are the average maximum unstable mass for rotating disc $M^{max}_{cl}$.
This scale mass has values $\ga10^8[M_\odot]$ for all our runs. We divided the mass range in
the bins $(5\times 10^5,10^6,5\times 10^6,10^7,5\times 10^7,10^8)[M_\odot]$. Our simulations 
form clumps in the range $M_{clump}\sim$ few $10^5$ to few $10^7[M_\odot]$The NoSNe simulation
reaches the higher mass clump with $M_{clump}\approx 8\times 10^7[M_\odot]$ due to its
null feedback. Our SNe runs reached lower maximum clump mass 
$M_{clump}\approx 2\times 10^7[M_\odot]$ due to the SNe heating and our SNe0.5 
experiment form more massive clumps compared with our SN5.0 run.}
\label{fig:scalemass}
\end{figure}

We applied the clump finder algorithm of \citet{Padoan+2007} to our galactic disc
inside a $\sim1-1.5[kpc]$ box. The clump finder algorithm scans regions of density 
above $5\times 10^2n_{avg}$, with $n_{avg}$ the average density inside the analyzed 
box which is of order $\sim 5[cm^{-3}]$. In practice it means that we look for gas 
clumps at densities above $\sim 10^3[cm^{-3}]$. The scan is performed increasing the 
density by a fraction $\delta n/n=0.25$ until the maximum box density is reached. For 
each step the algorithm selects the over-densities with masses above the Bonnor-Ebert 
mass in order to define a gravitationally unstable gas clump. This algorithm gave us 
clump masses in the range $\sim$ few $10^5-10^8[M_\odot]$. Figure \ref{fig:scalemass} 
shows the clump mass function found in each of our simulations at different redshifts. 
In order to complement this analysis we have computed the mass associated to the 
maximum unstable scale length of a rotating disc \citep{Escala&Larson08}
\begin{equation}
M_{cl}^{max}=\frac{\pi^4 G^2 \Sigma_{gas}^3}{4\Omega^4}.
\end{equation}

The vertical lines of figure \ref{fig:scalemass} mark the average $M_{cl}^{max}$ at each 
sampled redshift. Our NoSNe run formed the bigger gas clumps. In this case due to the
lack of feedback the most massive objects ($M_{clump}\approx8\times10^7 [M_\odot]$) 
can survive at different redshifts. Such mass is of the order of the expected 
$M_{cl}^{max}\ga10^8[M_\odot]$.
  
The SNe runs could form objects as big as $M_{clump}\approx2\times10^7 [M_\odot]$. These 
masses are below the $M_{cl}^{max}\ga$ few $10^8[M_\odot]$. The SNe0.5 simulation
forms much more massive objects compared with the SNe5.0 run. Due to the extreme
feedback of the SNe5.0 experiment it is not easy for the clump to survive in such a violent
environment. This is why the SNe5.0 simulation forms less clumps throughout its evolution. 

All the clumps formed in our simulations have sizes in the range of $\lambda_{clump}\sim$
few $10^0 [pc]$ to few $\sim 10^1 [pc]$ (note that as this size is associated to all the cells 
above the threshold $n_{avg}$, then it is a minimum size because it could increase if we 
reduce $n_{avg}$). These sizes are below the unstable length scale (averaged on the inner 
$\sim 1[kpc]$) associated to the maximum clump mass: 
$\lambda_{clump}< \lambda_{cl}^{max}=4\pi^2 G \Sigma_{gas}/\Omega^2\sim$ few $10^2 [pc]$.

\subsection{Mass transport on the disc}
It is well known that in a cosmological context the large scale ($\ga R_{\rm vir}$) gas 
cooling flows associated to DM filaments converging onto DM haloes have influence on 
the small scales ($\la R_{\rm vir}$) galactic AM 
\citep[e.g. ][]{Powelletal2011,PrietoSpin,Danovich2015}. Such an interplay between large 
and small scales suggests that the mass/AM transport analysis should be performed 
taking into account both regimes. 

\subsubsection{Stresses on the disc}
The MT on the galactic disc can be studied based on the momentum conservation equation. 
Written in its conservative form this equation tell us that the local variation of momentum is 
due to the rate of momentum fluxes:
\begin{equation}
\frac{\partial (\rho v_i)}{\partial t}+\frac{\partial}{\partial x_k}(R_{ik}+P_{ik}-G_{ik})=0,
\label{peq}
\end{equation}
where $\rho$ is the gas density, $x_i$ are the Cartesian coordinates and $v_i$ are the 
Cartesian components of the gas velocity. All the terms inside the divergence are related
with the rates of momentum flux and they can be written as follow:
\begin{equation}
R_{ik}=\rho v_i v_k.
\end{equation}
is the term associated to the Reynolds (or hydrodynamic) stress. It is the momentum
flux term associated to the total fluid movement. Instead of being a momentum 
flux source it quantifies the transported momentum due to the addition of different 
phenomena on the disc, namely gravitational stresses, magnetic stresses, viscous stresses 
or pressure stresses.

\begin{equation}
P_{ik}=\delta_{ik}P.
\end{equation}
This is the pressure term, where $\delta_{ik}$ is the Kronecker delta symbol, $P$ is the gas 
pressure and its gradient will be a source of torque as we will show in the following lines.

\begin{equation}
G_{ik}=\frac{1}{4\pi G}\left[\frac{\partial \phi}{\partial x_i}\frac{\partial \phi}{\partial x_k}-\frac{1}{2}(\nabla\phi)^2\delta_{ik}\right].
\end{equation}
with $\phi$ the gravitational potential and $G$ Newton's constant. $G_{ik}$ 
is the term associated to the gravitational stress and it is related with the 
movements of the fluid due to the gravitational acceleration. This term also will be 
a source of torques acting on the fluid as we will show later.

Because we are not including magnetic fields we have neglected the term associated to it.
Furthermore, the dissipative-viscous term is negligible in this context and will not 
be taken into account in the following discussion \citep[e.g. ][]{Balbus2003}.

In the disc MT context it is useful to quantify the momentum transport in the 
$\hat{r}$ direction due to processes in the $\hat{\theta}$ direction where $\hat{r}$ 
and $\hat{\theta}$ are the radial and the azimuthal cylindrical coordinates,
respectively. If $F_{r\theta}$ is the rate of momentum flux in the $\hat{r}$ direction
due to the processes in the $\hat{\theta}$ direction associated to any of the stresses
mentioned above, in general we can write (see appendix \ref{appendixA})
\begin{equation}
F_{r\theta}=\frac{1}{2}(F_{yy}-F_{xx})\sin2\theta+F_{xy}\cos2\theta.
\end{equation}

After some algebra it is possible to write the momentum fluxes for each of our sources
as follow \citep[e.g. ][ and references therein]{Balbus2003,Fromang+2004}:

\begin{equation}
R_{r\theta}=\rho v_r v_\theta,
\end{equation}

\begin{equation}
P_{r\theta}=0 \quad {\rm and}
\end{equation}


\begin{equation}
G_{r\theta}=\frac{1}{4\pi G}\nabla_r\phi \nabla_\theta \phi,
\end{equation}

It is worth noticing that in the case of $\theta$ symmetry the gravitational term 
vanishes. In other words, any density perturbation in the $\hat{\theta}$ direction, 
e.g. an asymmetric density distribution of gas clumps in the disc, will cause a momentum flux in 
the $\hat{r}$ direction. This will be the term associated to the VDI as we will show
later.

The terms associated to the Reynolds and the gravitational stress as defined in 
the above expressions are averaged in space in order to quantify the radial 
momentum flux associated to perturbations in the azimuthal direction 
\citep{Hawley2000} The Reynolds and the gravitational stress are defined as
follow\footnote{An alternative definition of the Reynolds stress from \citet{
Hawley2000} is presented in appendix \ref{appendixF} with similar results.}:

\begin{equation}
\langle R_{r\theta}\rangle=\langle \rho v_r \delta v_\theta\rangle,
\label{reynoldsstress}
\end{equation}

\begin{equation}
\langle G_{r\theta}\rangle=\frac{1}{4\pi G}\langle\nabla_r\phi\nabla_\theta\phi\rangle,
\end{equation}
where $\delta v_\theta\equiv v_\theta-\langle v_\theta\rangle$, 
$\langle v_\theta\rangle$ the average circular velocity of the fluid and the
averages are computed as
\begin{equation}
\langle f(r,z,\theta)\rangle=\frac{\int\int r d\theta dz f(r,z,\theta)\rho}{\int\int r d\theta dz\rho}.
\end{equation} 

In this context it is useful to define an $\alpha$ parameter for each of our stresses.
For a given rate of momentum flux, following \citet{Gammie2001}, we define:
\begin{equation}
\alpha_{r\theta}=\alpha_{R,r\theta}+\alpha_{G,r\theta}=\Big\langle\frac{R_{r\theta}+G_{r\theta}}{P}\Big\rangle
\end{equation}
Each $\alpha$ parameter is interpreted as the rate of momentum flux associated with a 
given process normalized by the gas pressure. Because the gas pressure is 
$P\sim (\rho c_s)\times c_s$
it can be interpreted as a ``thermal momentum'' advected at the sound speed or
as a ``thermal rate of momentum flux''. In this sense an $\alpha \ga 1$ is a sign of 
super-sonic movements in the fluid. This parameter is $\alpha\approx 0.02$ for ionized 
and magnetized discs \citep{Fromang+2004,NelsenPapaloizou2003} and observations of
proto-stellar accretion discs \citep{Hartmann+1998} and optical variability of AGN 
\citep{Starling+2004} give an alpha parameter $\sim 0.01$. Due to the turbulent 
(i.e. high Mach number, see appendix \ref{appendixD}) nature of the environments studied 
here, the alphas will typically be higher than 1. In fact, $\alpha_R\la M^2$ with
$M$ the gas Mach number.

Figure \ref{fig:stressprof} shows the radial values of $\alpha_{R,r\theta}$ in the 
top row and $\alpha_{G,r\theta}$ in the bottom row for our three simulations in different
columns. 

The first thing that we should notice from this figure is that the Reynolds $\alpha$ parameters 
are not constant neither in time nor in space and furthermore they reach values well above unity. 
In other words, our high redshift galactic discs are not in a steady state. Such a dynamical 
condition does not allow use of the \citet{SS1973} mass accretion rate expression as a function 
of the computed $\alpha$ parameter. (See appendix \ref{appendixB} for a more detailed discussion.) 
Instead of that we must compute a mass accretion rate directly from our data.

From figure \ref{fig:stressprof} it is clear that the Reynolds stress tends to be much 
larger than the gravitational stress and then it dominates the MT process in 
most of the cases (note that both top and bottom panels are not in the same $\alpha$ 
range). In other words, the rate of momentum flux associated to the gravitational 
potential gradients is lower than the rate of momentum flux associated to local 
turbulent motions of the gas in most of the cases. Such high values of $\alpha_R$ 
are associated to high velocity dispersions which can be an order of magnitude above 
the sound speed. We note that our two SNe runs have lower $\alpha_R$ due to the higher 
sound speed in their environment.

Here we emphasize that the Reynolds tensor is not a source of momentum flux, in the 
sense that if we start the disc evolution from a spherical circular rotation state, 
i.e. without a radial velocity component, with null viscosity and one small
gravitational potential perturbation in the $\hat{\theta}$ direction the variation 
in momentum will be associated to the gravitational stress and the appearance of 
the Reynolds stress will be a consequence of this process. 

It is interesting to note that the $\alpha_{G,r\theta}$ parameter in our NoSNe and SNe0.5
has a decreasing trend with the galactic radius at some redshifts: the smaller the 
radius the larger the gravitational stresses. If we take into account that the 
accreted material tends to concentrate in the inner part of the galaxy then it 
is reasonable that the larger gravitational stresses act at small radii. In the 
NoSNe run it is of the order of the pressure at the galactic center at all redshifts,
whereas in the SNe0.5 run it is comparable to the pressure at high z. Due to the
high feedback the SNe5.0 run is dominated by the Reynolds stress in all the sampled
redshifts.

\subsubsection{Torques on the disc}

After observing that the Reynolds stress associated to the gas turbulent motions 
dominates the rate of momentum flux in the disc and that the gravitational $\alpha$
tends to reach its maximum at the central galactic region, it is relevant for the 
MT study to analyze the torques acting in the disc associated to forces in the 
$\hat{\theta}$ direction. In order to do that we compute the torques associated to 
both the gravity and the gas pressure for our systems. We define these two quantities 
as:
\begin{equation}
\vec{\tau}_G=\vec{r}\times\nabla\phi,
\end{equation}
\begin{equation}
\vec{\tau}_P=\vec{r}\times\frac{\nabla P}{\rho}
\end{equation}
which actually are specific torques, i.e. torques per unit gas mass. These two 
terms will act as a source of AM transport in the galactic disc and will give us
some clues about the MT process in high redshift galactic discs. In order to compute
this we have defined the radial origin to be in the cell where the sink particle is 
set.

Figure \ref{fig:profTorque} show the ratio between $\tau_G$ and 
$\tau_P$, with $\tau_i\equiv|\vec{\tau}_i|$. The NoSNe run shows a decreasing 
trend with radius, like in the $\alpha_{G}$ profile. The pressure gradients 
tend to dominate above $\sim 100 [pc]$ and the gravity force dominates in the innermost 
region. As already shown in the alpha profiles, in the SNe0.5 run the gravity dominates 
the central part of the system at high $z$ and at lower redshifts the pressure
torques are the source of mass transport. And finally, due to the high feedback
which is able to create strong shocks and destroy gas clumps, the SNe5.0 simulation
tends to be dominated by torques associated to pressure gradients in line
with the previous alpha results.

\begin{figure*}
\centering
\includegraphics[width=2.0\columnwidth,height=1.0\columnwidth]{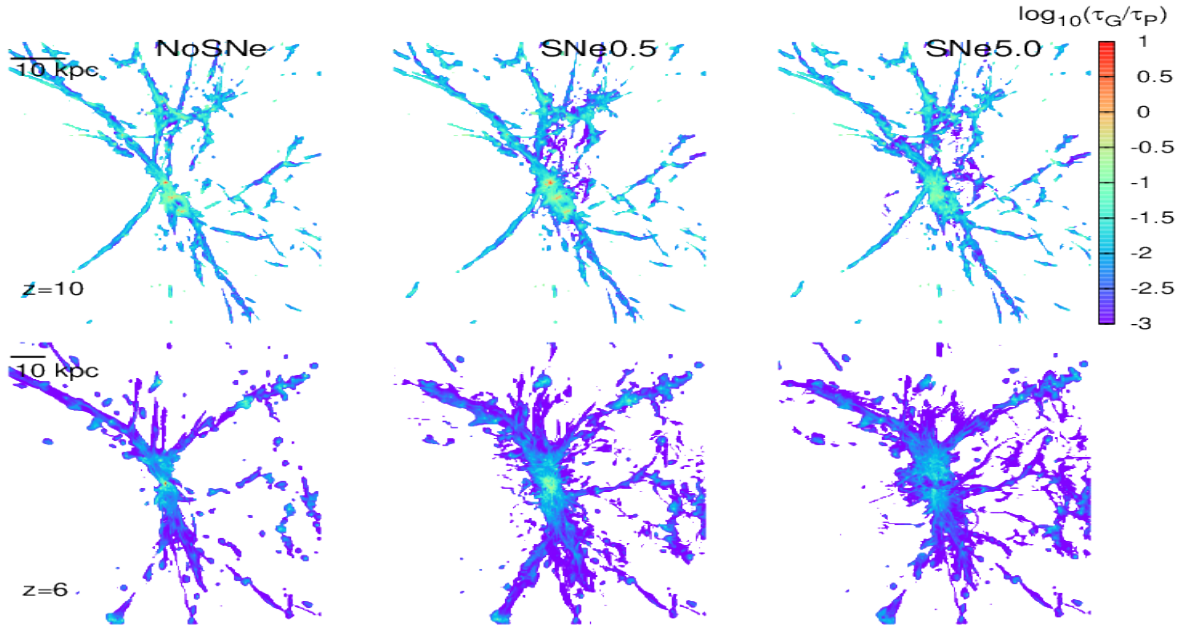}
\caption{Modulus of the mass weighted gravitational to pressure gradients torque 
ratio at $z=10$ in the top row and at $z=6$ in the bottom row. From left to right:
NoSNe, SNe0.5 and SNe5.0. It is interesting that the pressure torque dominates
over the gravitational torque in most of the mapped filamentary dense regions. The
gravitational torque increases its influence at the central region of filaments 
and around gas over-densities. Such a fact confirms our previous finding based on the
torques ratio radial profiles: gravitational torque increases its influence in the
central galactic region. Such a behavior is not true in our SNe runs where the SNe feedback
creates a region dominated by pressure gradients at the galactic center.}
\label{fig:multiTorque}
\end{figure*}

As a complement to our findings it is useful to take a look at torques at large scales.
Figure \ref{fig:multiTorque} shows the ratio of the total torques (not only the $\hat{z}$
component in the disc) $\tau_G/\tau_P$ for our three runs at two different redshifts, at
$z=10$ in the top row and at $z=6$ in the bottom row.

The maps take into account the gas with density above $50\times\Omega_b\rho_c$, where $\rho_c$ 
is the critical density of the Universe. Such a cut in density was set by inspection 
in order to have a clear view of the filaments around the central DM halo. 

It is interesting to note that in our three simulations the border of the filaments is 
clearly dominated by the pressure torque: material from voids falls onto the filamentary 
DM structure creating large pressure gradients \citep{Pichon+2011,Danovich2015}.
There it loses part of its angular momentum and flows onto the DM halo. 

At high redshift it is possible to see that inside the filaments the gravitational 
torque is $\la 0.1$ of the pressure torque. The picture changes when we look 
at the bottom panels, there the shocked filaments have a ratio $\tau_G/\tau_P\la 10^{-2}$. 

\begin{figure}
\centering
\includegraphics[width=1.0\columnwidth]{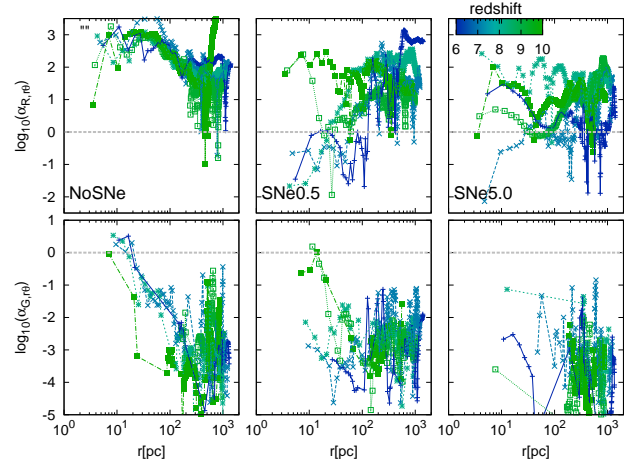}
\caption{The $\alpha_{r\theta}$ parameters in the disc as a function of radius for 
different redshifts. The gray dashed line marks the $\alpha=1$ position. From left to right:
NoSNe, SNe0.5 and SNe5.0. The top row shows the hydrodynamic $\alpha$ and 
the bottom row the gravitational $\alpha$. The $\alpha_R$ are much larger than one in most 
of our cases. Due to the high Mach number of the NoSNe run it shows higher values. Due to 
the high temperatures reached after SNe explosions our SNe runs have lower Reynolds alphas. 
The gravitational $\alpha$ is much lower than the Reynolds stress. It reaches higher values 
at the central galactic region for NoSNe and SNe0.5 ($z\ga 9$). The SNe5.0 run shows 
lower values due to the destruction of dense gas features and higher gas temperatures.}
\label{fig:stressprof}
\end{figure}

It is possible to find regions with a ratio $\tau_G/\tau_P> 10^{-2}$ around gas 
over-densities and near the main central halo. All the gas over-densities, in general 
associated to DM haloes at these scales, have a higher gravitational to pressure 
torque ratio. In particular at $z=6$ we can see that the central galactic region 
for the NoSNe simulation is dominated by the gravitational torque. This is not
the case for both our SNe runs where at low redshift the pressure torque is dominating
the AM re-distribution. Such behavior confirm the radial profile results of figure 
\ref{fig:profTorque} and the $\alpha$ parameters of figure \ref{fig:stressprof}. 

At the edge of the galactic disc the pressure torque associated to the in-falling 
shocked material tends to dominate AM variations whereas at the central region the 
gravitational potential gradient is the main source of torque in the NoSNe simulation. 
In the SNe runs the energy injection spread out the high density material and there is 
a more flat potential at the center of the galaxy implying a non clear gravity domination 
there. Such behavior is more evident in our SNe5.0 run where it is possible to see a dark 
region in the center of the system.

Having clarified that the source of pressure gradients are the shocks associated to both the
filamentary incoming material from the cosmic web and the SNe explosions, it is 
interesting to elucidate the origin of the gravitational torque acting mainly in the 
central region of the galactic disc. In order to do that it is useful to study the 
density distribution in the disc. In particular, it is worth computing the Fourier 
modes associated to the gas mass surface density:
\begin{equation}
c_m=\frac{1}{2\pi}\int_{-\pi}^{\pi}d\theta\int_{0}^{\infty}dr e^{im\theta} r\Sigma(r,\theta).
\end{equation}

\begin{figure}
\centering
\includegraphics[width=1.0\columnwidth]{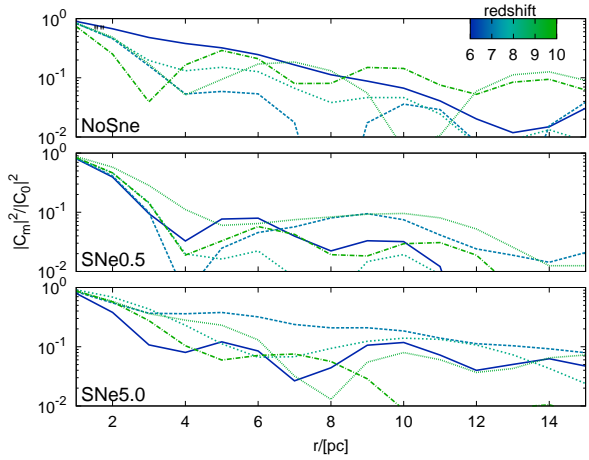}
\caption{The Fourier modes associated to the mass surface density spectrum on the disc. 
From top to bottom: NoSNe, SNe0.5 and SNe5.0. Despite of the $m=1$ 
mode has the higher power, the mode $m=2$ is also important being a fraction $\ga0.5$
of the $|c_1|^2$ at all redshifts. Furthermore, all the other $m>3$ modes have a 
contribution of roughly similar order between them. In other words, the disc 
have developed a complex azimuthal structure allowing gravitational torques on the
galaxy.}
\label{fig:cmode}
\end{figure}

Figure \ref{fig:cmode} shows the square of each Fourier mode (from $m=1$ to $m=15$) 
normalized by $|c_0|^2$ for five different redshifts for our three runs. It is clear 
from the figure that the $m=1$ mode has the highest power in the spectrum for all the 
shown redshifts. Despite that, it is also possible to see that the difference in 
power between the first and the second mode is not too much for all the sampled redshifts, 
i.e. $|c_2|^2/|c_1|^2\ga 0.5$. In this sense it is not possible to say that the first 
mode is the only contribution to the surface density spectrum because the second mode 
(and even the third one) is also important. Furthermore, it is worth noticing that 
the powers with $m>2$ are also there and they have values $\ga 10^{-2}$ below $m\approx 6$. 
It is interesting to compare our result with the one shown in \citet{Krumholz+2007} where
they found that the source of the torques on a proto-stellar disc is associated to the 
domination of the $m=1$ mode due to the SLING instability \citep{Adams1989,Shu1990}. 
In their case the first power is at least one order of magnitude higher than the $m=2$ 
mode with an increasing domination of $m=1$ mode with time. They argue that the $m=1$ spiral 
mode produces global torques which are able to efficiently transport AM. In our case, 
the global perturbation will be associated to a more complex disc structure. The 
reason for this difference will be clear after looking at the surface gas density 
projections. 

Figure \ref{fig:sigmagas} reflects the fact that the power spectrum of the gas 
surface density shows power for different modes $m$. There we can see a complex
spiral-clumpy structure defining the galactic disc. Such density field features 
create an inhomogeneous gravitational potential field which will exert torques on 
the surrounding media. In particular, the clumps formed on the disc by gravitational 
instabilities interact between themselves migrating to the central galactic region: 
the VDI acts on these high redshift clumpy galactic discs \citep{Bournaud+2007}. 

It is worth noticing that due to the SNe energy injection in the SNe runs the disc 
takes a longer time to appear comparable to the NoSne experiment. Whereas the NoSne 
simulation develops a disc that is progressively disturbed in time by no more than mergers, 
the SNe runs show a disturbed clumpy environment characteristic of turbulent gas where
the SNe explosions disrupt the galaxy with a strong effect on the central BH accretion
rate as we will see in the next sub section. Furthermore, due to the metal release
in the SNe runs the gas can cool more efficiently than in the NoSNe simulation. Such 
an important difference allows the gas to form more self gravitating over-densities 
and produce the clumpy galaxies shown at redshift $z\la8$ in the second and third 
columns of figure \ref{fig:sigmagas}.

\subsubsection{Mass accretion and BH growth}
High redshift galaxies are far from isolated systems. In fact, as has been shown above, 
they are very dynamic, in the sense that they are being built up in environments 
disturbed by filamentary accretion from the cosmic web, mergers 
and SNe explosions which affect the transport of AM. In the context 
of BH evolution at high redshift it is relevant to study and quantify the mass 
accretion rate in the galactic disc due to the processes described in the previous 
sub section, and more relevant yet is the quantification of the mass accretion rate onto
the central BH and the relation of its mass accretion with the large scale filamentary
inflows. 

\begin{figure}
\centering
\includegraphics[width=1.0\columnwidth]{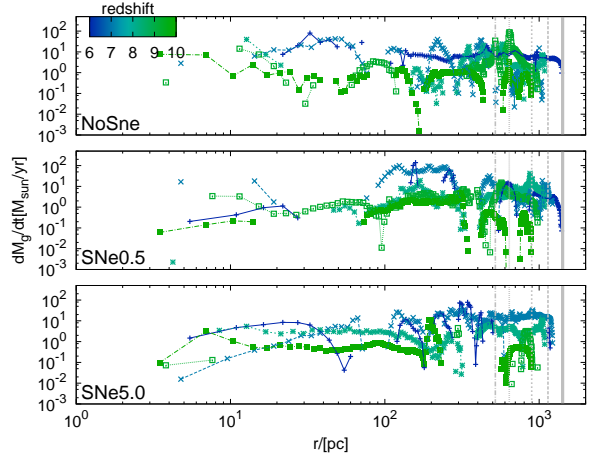}
\caption{Mass accretion rate radial profiles for our three runs. The vertical lines 
mark $0.1R_{vir}$ at each redshift. In all simulations the accretion
rate has huge fluctuations between $\sim$ few $10^{-2}$ and $\sim$ few $10^{1}[M_\odot/yr]$, 
another proof of the highly dynamic nature of the system. The SNe runs show 
a less continuous accretion with lower values. In particular, our SNe simulations
have a $\sim$ few $1[M_\odot/yr]$ at the end of the simulation.}
\label{fig:dotMprof}
\end{figure}

Figure \ref{fig:dotMprof} shows the radial gas mass accretion rate on the disc 
as a function of radius inside $\sim 0.1R_{vir}$ at different redshifts for our 
three runs. We defined the mass accretion on the disc as:
\begin{equation}
\frac{dM_{g}}{dt}=-2\pi r\Sigma_g v_r.
\end{equation}
The radial coordinate $r$ is defined in the disc plane and the gas SD $\Sigma_g$ and 
the radial velocity $v_r$ are cylindrical shell averages in the $\hat{z}$ direction.

We note that the NoSNe simulation shows continious lines 
at almost all redshifts and radii, but our SNe simulations present non
continuous lines due to gas outflows. Such features are another proof of 
the highly dynamic environment where the first galactic discs are formed. 
The mass accretion rate fluctuates roughly between 
$\sim10^{-2}$ and $\sim10^{2}[M_\odot/yr]$ in the range of radius shown in the 
figure. As mentioned above such a huge dispersion reflects the fluctuating 
conditions of the galactic disc environment at high $z$, where due to the continuous 
gas injection through filamentary accretion the evolution is far from the secular 
type we see in low redshift galaxies.  

At $z=6$ the NoSNe run has an accretion rate of $\dot{M}_g\approx7[M_\odot/yr]$ on
the disc. This value is higher than the $\dot{M}_g\approx4[M_\odot/yr]$ in the case
of SNe0.5 and $\dot{M}_g\approx3[M_\odot/yr]$ for the SNe5.0 simulation. This trend
can be related to the SNe feedback strength. As we will see below, the SNe feedback
will have important effects on the BH growth and on the mass accretion at larger
scales also.

\begin{figure}
\centering
\includegraphics[width=1.0\columnwidth]{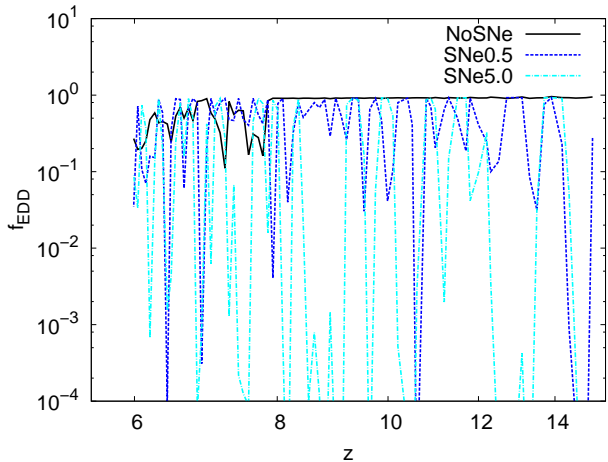}
\caption{BH mass accretion rate normalized by the Eddington accretion, 
$f_{EDD}$ as a solid black line for the NoSNe run, as a dashed blue line for the SNe0.5 
run and as a dot-dashed line for the SNe5.0 run. From the figure the increasing feedback 
perturbation on the BH growth is clear. It has an average value of 
$\langle f_{EDD}\rangle\approx0.75$ in the NoSne case and 
$\langle f_{EDD}\rangle\approx0.5$ in our SNe runs throughout the BH evolution.}
\label{fig:accrate}
\end{figure}

Figure \ref{fig:accrate} shows the ratio of the BH mass accretion rate to the Eddington
mass accretion rate $f_{EDD}\equiv\dot{M}_{BH}/\dot{M}_{EDD}$. The NoSNe
run BH accretes matter at the Eddington limit until $z\approx8$. At this redshift
the central galactic region undergoes several mergers losing a lot of gas, 
leaving the BH with almost no material to consume. After this event the accretion 
rate fluctuates until the end of the simulation. Our SNe run has an Eddington limited
BH accretion rate $\langle f_{EDD}\rangle\approx 0.75$ throughout its evolution and 
a $\langle f_{EDD}\rangle\approx 0.5$ below $z=8$.

From figure \ref{fig:accrate} the effect of SNe feedback on the BH growth is clear.
SNe feedback perturbs the BH accretion rate from the beginning of its evolution decreasing 
its value until $f_{EDD}\la10^{-4}$ in the SNe0.5 run and reaching even lower values in the 
SNe5.0 simulation. Such a difference in the BH accretion rate is translated into a 
$\langle f_{EDD}\rangle\approx 0.5$ for SNe0.5 throughout its evolution and a
$\langle f_{EDD}\rangle\approx 0.3$ for the SNe5.0 experiment. These values do not
change significantly when we average below $z=8$. The mass accretion rate onto the 
BH at the end of the simulations is $\dot{M}_{BH}\approx 8[M_\odot/yr]$, 
$\dot{M}_{BH}\approx 0.03[M_\odot/yr]$ and $\dot{M}_{BH}\approx 0.003[M_\odot/yr]$ for the
NoSNe, SNe0.5 and SNe5.0 run, respectively (see appendix \ref{appendixE}).

\begin{figure}
\centering
\includegraphics[width=1.0\columnwidth]{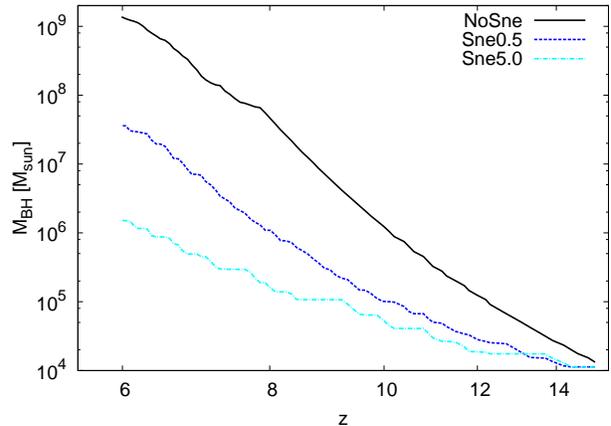}
\caption{BH mass evolution for our three simulations: NoSne (solid black line), SNe0.5 
(dashed blue line) and SNe5.0 (dot-dashed cyan line). The NoSNe BH reaches a mass of 
$1.4\times 10^{9}M_\odot$ at the end of the simulation. Such a high mass was reached because 
most of the time the BH was accreting at the Eddington limit. In the SNe0.5 
run the sink particle reaches a final mass of $3.6\times10^7M_\odot$ and our SNe5.0 
BH mass reaches $1.5\times10^6M_\odot$ due to the extreme feedback.}
\label{fig:sinkmass}
\end{figure}

Figure \ref{fig:sinkmass} shows the BH mass evolution as a function of redshift. 
From this figure it is possible to see the effect of the different behavior in the 
accretion rate, as shown in figure \ref{fig:accrate}. Whereas the NoSne sink has an 
approximately exponential evolution ending with a mass $M_{BH}=1.4\times10^9[M_\odot]$, 
due to the SNe feedback our SNe runs show episodes of no growth at some redshift.
Such a feature is much clearer in our SNe5.0 run (see between $z=14-12$ or 
$z\approx 9$ for example). The final mass in these two runs was $M_{BH}=3.6\times10^7[M_\odot]$
for SNe0.5 and $M_{BH}=1.5\times10^6[M_\odot]$ for SNe5.0.

\subsection{Mass transport on larger scales}
At high redshift we cannot study the small-scale galactic phenomena without taking
into account the effects of the large scale structure in a cosmological context. Here 
we study the behavior of the mass accretion rate above the $\sim$ kpc scales, i.e. beyond 
the galactic disc edge.

\begin{figure}
\centering
\includegraphics[width=1.0\columnwidth]{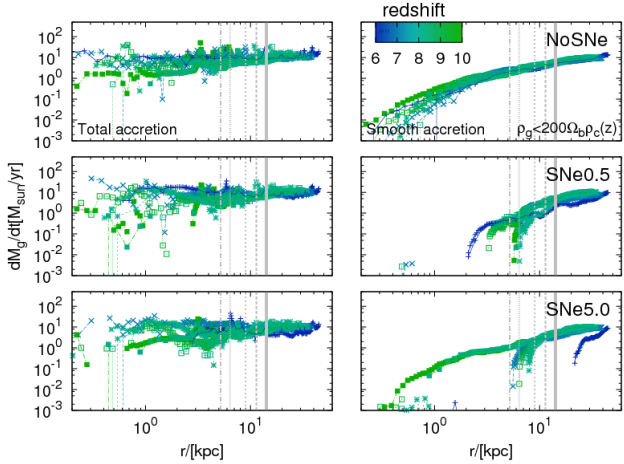}
\caption{Left column: Same as figure \ref{fig:dotMprof} but for larger radii taking into 
account material till $\sim 3R_{vir}$ around the central halo. From top to bottom:
NoSNe, SNe0.5 and SNe5.0. Right column: Same as left column but for 
the smooth accretion, i.e. for gas with a density below the collapse density 
$\rho_g< 200\Omega_b\rho_c$. Beyond the virial radius the total accretion has 
a floor similar to the smooth accretion. Inside the virial radius the accretion
rate is dominated by dense gas. The SNe explosions have a clear effect on the 
smooth accretion. At $\sim$ kpc scales the smooth accretion is practically erased due 
to the SNe heating.}
\label{fig:accrateLS}
\end{figure}

Figure \ref{fig:accrateLS} shows the mass accretion rate out to $\sim 3R_{vir}$. The 
mass accretion has been computed taking into account all the mass crossing a spherical 
shell at a given radius centered at the sink cell position:
\begin{equation}
\frac{dM_{g}}{dt}=-4\pi r^2\rho v_r.
\end{equation}

The left column of figure \ref{fig:accrateLS} shows the total mass accretion rate 
for our three simulations. In the right column we have plotted the mass accretion 
rate associated to gas densities below 
$\rho_{coll}=18\pi^2\Omega_b\rho_c\approx 200\Omega_b\rho_c$, 
with $\rho_c$ the critical density of the Universe. The vertical lines mark the DM
virial radius at each sampled redshift. 

The right column of the figure can be interpreted 
as smooth accretion associated to non collapsed objects. The NoSNe panel shows 
a smooth decreasing behavior almost independent of redshift above $\sim2[kpc]$. 
The smooth mass accretion rate presents roughly constant values above the virial 
radius, with accretion rates of the order $\sim10^1[M_\odot /yr]$. Such a value is
consistent with the one found by e.g. \citet{Dekel+2009} and \citet{Kimm+2015}
for a $\sim10^{10}[M_\odot]$ halo at high redshift \citep[see also ][]{Neistein+2006}. 
On the other hand, the SNe run panels have a more irregular decreasing behavior with 
a notable dependence on redshift due to SNe explosions. In these runs the SNe 
feedback is able to heat up the gas and create hot low density gas outflows
almost depleting the system of low density gas. In particular, for a number of 
redshifts it is possible to see that the smooth accretion is practically erased at 
radii $\la R_{vir}$, a clear signal of low density gas evaporation. In other words, 
due to SNe explosions only the dense gas is able to flow into the inner $\la1 [kpc]$ 
region of the galaxy. 

\begin{figure}
\centering
\includegraphics[width=1.0\columnwidth]{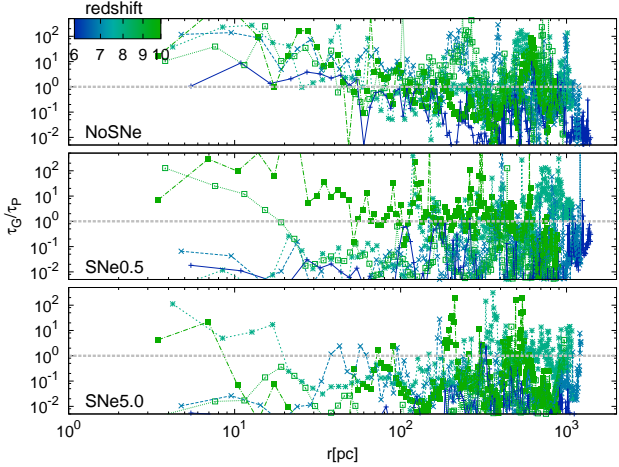}
\caption{Gravitational torque to pressure gradient torque ratio as a function of radius
for different redshifts. From top to bottom: NoSNe, SNe0.5 and SNe5.0. The gray dashed line 
marks the $\tau_G/\tau_P=1$ state. The NoSNe run MT tends to be dominated by gravitational
torque in the inner galactic region $r\la 100 [pc]$ at all redshifts. Beyond that radius the
pressure gradients and gravity work together to re-distribute AM. The SNe0.5 run tends to
be dominated by pressure with a central gravity domination at high $z$, inside 
$r\la10-100 [pc]$. The pressure gradient domination is more clear in our SNe5.0 due to 
the extreme SNe feedback.}
\label{fig:profTorque}
\end{figure}

\subsection{Gas-stars-DM spin alignment}
\begin{figure}
\centering
\includegraphics[width=1.0\columnwidth]{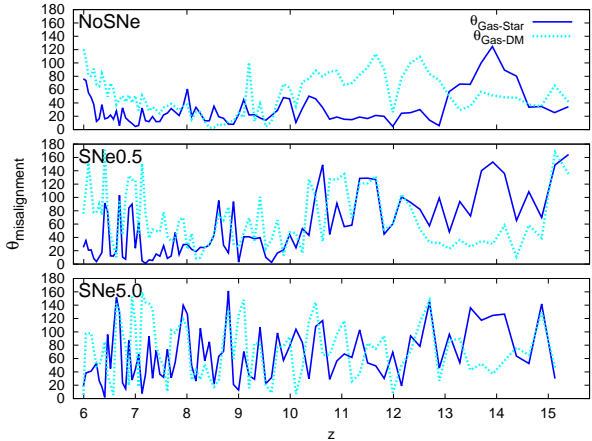}
\caption{The misalignment angle between the gas AM and stellar AM (solid blue line), and 
the gas AM and DM (short-dashed cyan line). From top to bottom: NoSNe, SNe0.5 and SNe5.0. 
The gas and DM show a fluctuating misalignment angle with a high value at the end of 
the simulation. Due to the collisional nature of the gas it decouple from the DM once 
the pressure torques start to work on it. For the same reason the SNe simulations have a
larger misalignment through out the simulation.}
\label{fig:spinangle}
\end{figure}

As a complementary analysis it is interesting to study the alignment between the AM
of the different components of the system, namely DM, gas and stars. Figure 
\ref{fig:spinangle} shows the alignment between the AM of the different components 
of our systems. The misalignment angle between the gas AM $\vec{l}_{Gas}$ 
and the component $i$ of the system $\vec{l}_{i}$ was computed as:
\begin{equation}
\cos(\theta_{Gas-i})=\frac{\vec{l}_{Gas}\cdot\vec{l}_{i}}{|\vec{l}_{Gas}||\vec{l}_i|}
\end{equation}

The rotational center to compute the gas, DM and stars AM was set at the sink position, 
$\vec{r}_{sink}$. This point coincides with the gravitational potential minimum cell
within $\sim50[pc]$ around the sink particle.

We computed the AM of the different components $i$ as
\begin{equation}
\vec{l}_{i}=\sum_j \Delta m_{i,j}(\vec{r}_{i,j}-\vec{r}_{c})\times(\vec{v}_{i,j}-\vec{v}_{c}),
\end{equation} 
where the sum is calculated inside $0.1R_{vir}$ for each component\footnote{We have
not made any distinction regarding the cell gas temperature or between disc and
bulge stars.}. $\vec{r}_{c}$ is the the center of the cell where the sink particle is
located, $\vec{v}_{c}$ is the average gas velocity of all cells inside a radius of
$5\Delta x$ around the sink position and $\Delta m_{i}$ is the mass of our different 
quantities: $i=$ gas, stars and DM.

From the figure we can see that the gas and the DM spins are far from aligned. The 
misalignment angle between them fluctuates from a parallel alignment 
$\theta_{Gas-DM}\la 10^\circ$ to an almost anti-parallel configuration 
$\theta_{Gas-DM}\approx 120^\circ$ in our NoSNe experiment. In our SNe runs
the fluctuations are more dramatic due to SNe explosions. 
Such a non-correlation between the AM vector of these two components has been 
studied before in, e.g. \citet{PrietoSpin}. In their work the authors noticed that 
after the cosmological turn around the gas can decouple from the DM due to its 
collisional nature: while the DM can feel only the gravity the gas can also feel the 
gas pressure gradients. Such pressure gradients are responsible for an extra torque 
on the baryonic component and its AM vector deviates from the DM AM orientation. As 
already shown in figure \ref{fig:profTorque} the pressure gradients are not negligible 
inside the virial radius of our haloes. Such torques are able to change the orientation 
of the gas AM and then create a misalignment between gas and DM AM vectors. 

\begin{figure}
\centering
\includegraphics[width=0.95\columnwidth,height=1.7\columnwidth]{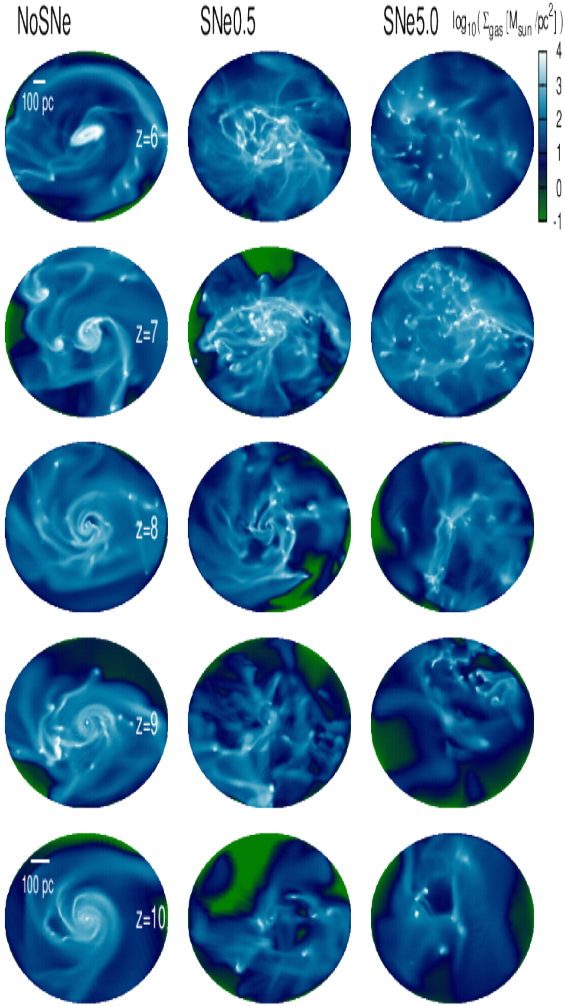}
\caption{Gas surface density projections for our runs at different $z$. 
From left to right: NoSNe, SNe0.5 and SNe5.0. The evolution of the density 
maps show that the galaxy develops a complex spiral clumpy structure supporting 
the existence of high $m$ powers in the Fourier analysis of figure \ref{fig:cmode}. 
Due to SNe explosion the spiral shape of the object appears only at $z\la8$ in the 
SNe0.5 run. Below this redshift the galaxy is successively destroyed by SNes and 
re-built by gravity. In our SNe5.0 run it is almost impossible to see a spiral shape
due to the extreme feedback.}
\label{fig:sigmagas}
\end{figure}

The alignment between gas and stars has a different behavior in 
our runs. For the NoSNe case it is possible to see that at high redshift, between 
$z\approx15$ and $z\approx13$, the stars and the gas had a very different spin
orientation. This is because at this stage the galaxy is starting to be built by
non spherical accretion and mergers, conditions which do not ensure an aligned 
configuration. After $z\approx13$ the gas and stars reach a rather similar spin 
orientation with a misalignment angle fluctuating around the value of 
$\theta_{Gas-Stars}\sim 20^\circ$. There the proto-galaxy can not be 
perturbed easily by minor mergers and acquires a defined spiral shape allowing the
gas-star alignment. Such an alignment is perturbed at redshift $z\approx10$. At 
this redshift the main DM halo suffers a number of minor mergers which can explain 
the spin angle perturbation. After that the gas and stars again reach an aligned
configuration which will be perturbed by mergers again at lower redshifts. 

The SNe runs show a much more perturbed gas-stars AM evolution. In this case, as well as
the merger perturbations the systems also feel the SNe explosions which continuously 
inject energy. The strong shocks associated to this phenomenon are able to decouple 
the gas AM from the stellar AM as we can see from the blue solid line. Such perturbations are
more common in our SNe5.0 run compared with our SNe0.5 run due to the stronger
feedback as we can see from figure \ref{fig:sigmagas} where the SNe5.0 simulation
shows a number of clumps instead of a defined spiral shape. 

\section{Discussion and Conclusions}
\label{conclusion}
By using cosmological hydrodynamic zoom-in simulations we have studied the 
MT process from $\sim$ few $10 [kpc]$ to $\sim$ few $1[pc]$ scales on a DM halo of 
$M\approx 3\times 10^{10}[M_\odot]$ at redshift $z=6$. We have studied the 
evolution of the system without SNe feedback (NoSNe run) with the delayed cooling
model for SNe feedback (SNe0.5 run) and with an extreme case of delayed cooling 
SNe feedback (SNe5.0 run).

We found that the SNe0.5 run is the best match with the D10 star burst galaxy sequence.
It covers about two decades in SD with the lowest scatter among our simulations. 
When we look at the stellar mass of the systems our SNe5.0 run shows a stellar
mass close to the expected value from B13, $M_\star/f_b M_{vir}\sim 10^{-2}$. 
Looking at this quantity our SNe0.5 run is still in the order of magnitude compared with B13 for a 
$\sim10^{10}[M_\odot]$ DM halo at high redshift. Such an offset can be related to the
``bursty'' nature of high redshift galaxies. In terms of the SFR, due to the extreme 
feedback, our SN5.0 run has the lowest values with a SFR $\sim 1[M_\odot/yr]$ at $z\la 8$. 
At the same $z$ range our SNe0.5 run has a SFR $\sim 10[M_\odot/yr]$ in agreement with
results from the W15 high $z$ galaxies. Despite this both SNe runs present low 
($\la 10^{-1}$) episodic SFR values due to the SNe heating. 

Our SNe experiments show lower gas fractions among our three simulations.
They have values $f_g\la0.85$ below $z=8$. If we look at the gas fraction below $z=7$ our
SNe0.5 run has the lowest value with $f_g\la0.8$ which is just within the upper limit 
for the SFR of the $z=7.5$ galaxy found by W15.

Following \citet{Gammie2001} we have computed the $\alpha$ parameters associated 
to both the Reynolds and the gravitational stresses. In other words, we have computed
both the Reynolds and the gravitational rate of momentum fluxes on the disc normalized
by the gas pressure. \citet{Gammie2001} showed that the $\alpha$ parameters
associated to radial mass transport are of order $\alpha\sim 10^{-2}$, reasonable values
for a subsonic stationary accretion disc. In our case the $\alpha$ parameters 
reach values above unity, meaning that the rate of momentum flux has values higher than 
the gas pressure $P=c_s\times(\rho c_s)$. Such high values are characteristic of a 
turbulent super-sonic environment associated to dynamical systems like the ones in our
simulations. The highly non-stationary gas behavior is confirmed also by the highly 
fluctuating values of $\alpha$ at all redshift.

We found that the Reynolds stress dominates over the gravitational one in most of the
analyzed redshifts. Here it is worth noting that the Reynolds stress tensor is a
measurement of the turbulent motions in the gas. In these systems the gas 
falls from large scales, channeled by filaments almost freely onto the DM halo 
central region, gaining super-sonic velocities. Through the virialization process
strong shocks are created developing a turbulent environment which is enhanced
due to SNe explosions. Under such conditions the rate of momentum flux associated to 
this term normalized by the gas pressure will be much higher than 1 if the rms gas 
velocity in the $\hat{r}$ and $\hat{\theta}$ directions are super-sonic. 

We emphasize that the Reynolds stress is not a source of mass transport but it is 
a measurement of the local rate of momentum transport triggered by other processes, 
namely pressure gradients, gravitational forces, magnetic fields or viscosity. 
In this sense, its high value simply tells us that throughout galaxy evolution 
there exists processes capable of transporting mass from large scales to small scales very
efficiently. In fact, in our systems, gravity triggers the mass flows through the
DM filamentary structure around the central halo and then a combined effect
of gravity and pressure gradients allows the MT in the disc. The Reynolds term tends 
to be higher for our NoSNe run where Mach numbers are higher due to the null SNe
heating.

The gravitational $\alpha$ parameter has a different behavior in our three experiments.
The NoSNe run shows a clear decreasing trend in radius until $r\sim 100 [pc]$ for all
sampled redshifts. Beyond that $\alpha_G$ fluctuates between values lower than 1. It 
reaches values $\sim 1$ at the central region. Such a behavior is telling us that
the gravitational term is more important at the central galactic region where matter
is more concentrated.  

Our SNe0.5 run has a peak above unity in the central region at high redshift decreasing
until $r\sim100 [pc]$. Beyond that radius it has a similar behavior compared with our
NoSNe simulation. Below redshift $z\sim 8$ the gravitational alpha parameter reduces its 
value to around $\sim10^{-3}$ with a lot of dispersion but always below $\sim10^{-1}$.
In this case the SNe feedback is able to deplete the central galactic region of gas
after $z\sim 8$ reducing the stresses associated to the gravitational gradients.

The SNe5.0 simulation has no peak at the galactic center. It seems to have the lowest
values at the central regions. Due to the extreme feedback adopted in this simulation 
it is much more difficult for gas to create dense structure producing important gravitational
forces. Furthermore, in this case the gas maintains higher temperatures implying 
higher pressures counteracting the gravitational effect.

The torques acting in the disc show the sources of angular momentum variations triggering
the MT process in these galaxies. In our systems the sources of torques are the pressure
gradient due to the shocks created through the virialization process and SNe explosions,
and the gravitational forces associated to gas inhomogeneities. 

As in the $\alpha_G$ analysis when we compute the gravitational to pressure gradient
ratio our NoSNe run shows a decreasing trend in radius. Gravity dominates over pressure 
and has a maximum at the central galactic regions reaching values $\sim 1$ at radius 
$\sim 100 [pc]$. Beyond this radius the pressure gradients tend to dominate the AM 
re-distribution. Without SNe feedback the pressure domination at large radius is associated
to shocks created by the large scale in-falling material to the central region of the host DM
halo. Despite the domination of the pressure gradients in the outer regions the system 
shows a number of regions where gravity acts showing a mixed contribution for the MT process.

In our SNe runs the domination of gravity at the central regions is not as clear as in the 
NoSNe run. In the SNe0.5 simulation the gravitational gradients dominate above $z\sim 9$
inside $r\la100 [pc]$. At lower redshifts the pressure gradients clearly dominate the 
torques at the inner $\sim 100 [pc]$. Beyond that radius again it is possible to see
a mixed torque contribution to the MT. A similar scenario is shown in our SNe5.0 simulation.
In this case the gravity can dominate the very central regions ($r\la$ few $10 [pc]$)
at high redshift and the pressure gradients have a more clear domination at larger
radii, but there is still a mixed contribution to the AM re-distribution.

When we look at the large scales related with the filamentary structure around the 
central DM halo it is possible to see that pressure torques dominate over gravitational 
torque in filaments. The central region of the filaments show an enhanced gravitational 
contribution, but it is not enough to be dominant. These results are consistent with the 
picture in which the material filling the voids falls onto the filamentary over-densities 
where it is channeled to the central DM halo region \citep{Pichon+2011,Danovich2015} 
by gravity. Once the gas reaches the filaments it feels the pressure gradient on the 
edge of the filaments and it loses part of its AM. Then gravity acts and transport the 
mass almost radially inside the cold filaments to the central region of the DM halo. Such 
a process allows the gas to reach the galactic edge almost at free-fall. There the gas 
pressure acts reducing its initially high radial velocity and at the same time exerting 
torques allowing the MT. Throughout this process the gravitational torques also work in 
the galactic gas helping the MT process in the disc.

A Fourier analysis of the disc gas surface density field for our runs shows that the 
density power spectrum has a number of excited modes. Despite the $m=1$ and $m=2$ 
modes dominating the power spectrum, the other modes do exist and have roughly 
comparable values between them. Such features tell us that the gas SD develops 
a complex structure throughout its evolution. The information given by the 
Fourier analysis is confirmed by visual inspection. The galactic discs develop 
spiral arms and gas clumps which interact between them by gravity. The gas clumps 
are formed from the cold gas flowing from the cosmic web onto the central DM halo 
region. The high gas fraction (which is $f_g\ga 60\%$) and cold environment 
is a perfect place to produce a clumpy galactic disc. The interaction between gas clumps, 
spiral arms and merged DM haloes exert gravitational torques which are capable of 
transporting mass onto the galactic center in times comparable to the dynamical 
time of the system: this is the so-called VDI \citep{Mandelker+2014,Bournaud+2007}.

Due to the process described above, i.e. large scale gravitational collapse inducing
filamentary accretion onto the DM central region and both gravitational and pressure
torques acting in the galaxy, the mass can flow through the galactic disc and reach
the galactic center. The radial mass accretion rate inside $\sim 0.1R_{vir}$ has 
huge fluctuations with values in the range $\sim(10^{-2}-10^{1})[M_\odot/yr]$ for our SNe
runs, a clear proof of a non-stationary and highly dynamic environment.

The high mass accretion rate in the high gas fraction disc allows the central 
BH to grow at the Eddington limit most of the time for the NoSNe run whereas
in the SNe runs it is clearly affected by the SNe explosions showing an intermittent 
Eddington-limited accretion rate. Despite this it can increase its mass 
substantially throughout the simulation. The violent events, namely mergers 
(which can also trigger mass accretion torquing the gas in the disc) and SNe 
explosions are not enough to stop the BH growth. The $10^4[M_\odot]$ BH seed can 
evolve until $M_{BH}=1.4\times 10^9[M_\odot]$ in our NoSNe experiment, 
$M_{BH}=3.6\times10^7[M_\odot]$ in our SNe0.5 and $M_{BH}=1.5\times10^6[M_\odot]$ in
our SNe5.0.

When we look at the mass transport beyond the virial radius we find that the large 
scale $r\ga R_{vir}$ mass accretion rate has a floor of the order $\la10^1[M_\odot/yr]$
with peaks associated to gas inside DM haloes of $\sim$ few $10^1[M_\odot/yr]$
in all our runs \citep[consistent with ][]{Dekel+2009,Kimm+2015}. Inside the virial radius 
the smooth accretion decreases monotonically
reaching values $\sim 10^{-1}-10^{-2}[M_\odot/yr]$ in the galactic outer regions, 
i.e. $r\sim 0.1R_{vir}$. These values change dramatically when we look at our feedback
simulations. In these cases the SNe feedback practically depletes the galactic central
region of low density gas. Due to the strong feedback effect only dense gas is able to
reach the outer regions of the central galaxy. The mass accretion rate associated to dense
gas is of the order $\sim 10^{0}-10^{1}[M_\odot/yr]$ in our NoSNe systems and it is almost 
devoid of discontinuities. On the other hand, despite the SNe runs reaching similar
accretion rates in the disc, they do have discontinuities, i.e regions of zero accretion
rate, affecting the amount of gas reaching the outer galactic region.

At the end of the simulation our NoSNe run shows an accretion rate
$\dot{M}_{BH}\approx 8[M_\odot/yr]$ which is similar to the total mass accretion in the
disc. Contrarily, our SNe0.5 run ends with $\dot{M}_{BH}\approx 3\times 10^{-2}[M_\odot/yr]$ 
and our SNe5.0 run reaches $\dot{M}_{BH}\approx 3\times 10^{-3}[M_\odot/yr]$ at the end of 
the experiment showing how important are the SNe explosions to the BH accretion rate.  

The gas AM vector orientation fluctuates a lot with respect to the DM spin vector through
out the system evolution. The gas and DM start their evolution with spin vectors roughly
aligned but once the pressure gradients increase due to virialization shocks, mergers
\citep[e.g. ][]{PrietoSpin} and SNe explosions they decouple reaching an almost anti-parallel
orientation at some stages. The alignment between these two components is more clear
in our NoSNe run where the angle between them is $\theta\la 60^\circ$ below
$z\approx 13$. The picture changes when we look at our SNe simulations, there the effect
of SN feedback is capable of changing the alignment from $\sim 0^\circ$ to $\ga100^\circ$
in $\sim$ few $10[Myr]$. Such an effect is stronger in our SNe5.0 where the big
angle fluctuations are present throughout the entire system evolution.

The inclusion of AGN feedback in our simulations certainly could change both the galaxy
and the BH evolution. The strong energy release in the gas can increase the gas
temperature and may suppress the star formation changing SFR properties of those
objects. Furthermore, due to the outflows associated to BH feedback the gas may 
not reach the central galactic region as easily as in the simulations presented here. 
A more detailed study of mass transport on high redshift galaxies with AGN feedback is 
left for a future study in preparation.

To summarize: In a cosmological context galaxies are formed inside knots of 
the cosmic web surrounded by filaments. The gas flows from voids to the DM filaments 
from all directions. There the gas piles up in the filamentary structure and its 
pressure gradient cancels part of its angular momentum. The pressure torques 
dominate the filamentary structure whereas the gravitational torques have a 
non-dominant enhancement at the center of filaments. Part of the material inside 
the filaments formed by dense cold gas, flows into the DM central halo in almost 
free-fall due to the host DM halo gravitational attraction. The transported 
cold gas reaches the DM halo with a high radial component of the velocity producing 
strong pressure gradients at the edge of the galactic disc. The constant inflowing of 
cold gas creates a high gas fraction and cold environment at the inner $\sim 0.1R_{vir}$. 
Such conditions promote a low Toomre parameter in the disc and so it becomes 
gravitationally unstable, forming very efficiently gas clumps of masses in the range
$\sim10^{5-8}[M_\odot]$ which interact between them, with the galactic spiral arms and with 
merged DM haloes. Such a clumpy environment produces regions in the disc dominated by
gravity torques, in other words the gravitational torque due to the VDI acts as a
source of MT in high redshift galaxies. The other, dominant, source of torques in our 
system is the pressure gradients. It is produced by SNe explosions and virialization shocks 
complementing the gravitational MT effect on this $z=6$ galaxy. The mass accretion rate in 
the disc triggered by pressure gradients and gravity can reach peaks of 
$\sim$ few $10^1[M_\odot/yr]$ and average values of $\sim$ few $10^0[M_\odot/yr]$ 
allowing an efficient BH mass growth. 

\section*{Acknowledgments}
J.P. and A.E. acknowledges the anonymous referee for the invaluable comments to improve this work.
J.P. acknowledges the support from proyecto anillo de ciencia y tecnologia ACT1101. A.E.
acknowledges partial support from the Center of Excellence in Astrophysics and Associated 
Technologies (PFB06), FONDECYT Regular Grant 1130458. Powered@NLHPC: This research was 
partially supported by the supercomputing infrastructure of the NLHPC (ECM-02). The 
Geryon cluster at the Centro de AstroIngenieria UC was extensively used for the analysis 
calculations performed in this paper. The Anillo ACT-86, FONDEQUIP AIC-57 and QUIMAL 
130008 provided funding for several improvements to the Geryon cluster. J.P. acknowledges 
the valuable comments and discussion from Yohan Dubois and Muhammad Latif. J.P. and A.E. 
acknowledge to Marta Volonteri for her enlightening comments on this work.

\section*{Appendix 1}
\label{appendixA}

In order to compute the momentum flux in the $\hat{r}$ direction due to 
processes in the $\hat{\theta}$ direction we projected the tensor $F_{ik}$ 
in the $\hat{\theta}$ and then in $\hat{r}$ direction:
\begin{equation}
F_{r\theta}=F_{ik}(\hat{x}^i\cdot\hat{\theta})(\hat{x}^k\cdot\hat{r})
\end{equation}
\begin{equation}
F_{r\theta}=\left[F_{xk}(\hat{x}\cdot\hat{\theta})+F_{yk}(\hat{y}\cdot\hat{\theta})\right](\hat{x}^k\cdot\hat{r})
\end{equation}

\begin{eqnarray}
F_{r\theta}&=&[F_{xx}(\hat{x}\cdot\hat{\theta})(\hat{x}\cdot\hat{r})+F_{xy}(\hat{x}\cdot\hat{\theta})(\hat{y}\cdot\hat{r})+ \nonumber \\
           & & F_{yx}(\hat{y}\cdot\hat{\theta})(\hat{x}\cdot\hat{r})+F_{yy}(\hat{y}\cdot\hat{\theta})(\hat{y}\cdot\hat{r})]
\end{eqnarray}

\begin{eqnarray}
F_{r\theta}&=&[-F_{xx}\sin\theta\cos\theta-F_{xy}\sin\theta\sin\theta+ \nonumber \\
           & &  F_{yx}\cos\theta\cos\theta+F_{yy}\cos\theta\sin\theta]
\end{eqnarray}

\begin{equation}
F_{r\theta}=\frac{1}{2}(F_{yy}-F_{xx})\sin2\theta+F_{xy}\cos2\theta,
\end{equation}

We can do a similar exercise in order to compute the flux of $\hat{z}$ AM in the 
$\hat{r}$ direction due to stresses in the $\hat{\theta}$ direction, $L_{rz}$. 
In this case we project in the $\hat{r}$ direction the $\hat{z}$ component of the 
AM associated to the stresses in the $\hat{\theta}$ direction:

\begin{equation}
L_{rz}=\epsilon_{jmi}x_m F_{ik}(\hat{x}^j\cdot\hat{z})(\hat{x}^k\cdot\hat{r})
\end{equation}

\begin{equation}
L_{rz}=\epsilon_{zmi}x_m F_{ik}(\hat{x}^k\cdot\hat{r})
\end{equation}

\begin{equation}
L_{rz}=[\epsilon_{zmi}x_m F_{ix}(\hat{x}\cdot\hat{r})+\epsilon_{zmi}x_m F_{iy}(\hat{y}\cdot\hat{r})]
\end{equation}

\begin{eqnarray}
L_{rz}&=&[\epsilon_{zyx}y F_{xx}(\hat{x}\cdot\hat{r})+\epsilon_{zxy}x F_{yx}(\hat{x}\cdot\hat{r}) \nonumber \\ 
      & & \epsilon_{zyx}y F_{xy}(\hat{y}\cdot\hat{r})+\epsilon_{zxy}x F_{yy}(\hat{y}\cdot\hat{r})]
\end{eqnarray}

\begin{equation}
L_{rz}=[-y F_{xx}+x F_{yx}](\hat{x}\cdot\hat{r}) + [-y F_{xy}+x F_{yy}](\hat{y}\cdot\hat{r})
\end{equation}

\begin{equation}
L_{rz}=(x F_{yx}-y F_{xx})\cos\theta + (x F_{yy}-y F_{xy})\sin\theta
\end{equation}

\section*{Appendix 2}
\label{appendixB}

\citet{SS1973} studied the MT process in a viscous disc. In such a model the MT is 
due to the effect of local viscous stresses. This process can be quantified 
by a rate of momentum flux term. This term can be written as a function of 
the viscosity $\nu$ and the velocity shear as follows:
\begin{equation}
S_{ik}=\rho\nu\left(\frac{\partial v_i}{\partial x_k}+\frac{\partial v_k}{\partial x_i}-\frac{2}{3}\delta_{ik}\nabla\cdot \vec{v}\right),
\end{equation}
Assuming the strong constraint of a steady state for the momentum evolution, 
i.e. $\partial(\rho v_i)/\partial t=0$, we obtain
\begin{equation}
\frac{\partial}{\partial x_k}(R_{ik}-S_{ik})=0.
\end{equation}
In order to compute the MT in such a steady state, i.e. when the term inside the 
partial derivative is equal to a constant $C$, we can write the above equation as
\begin{equation}
R_{r\theta}=S_{r\theta}+C=\frac{3}{2}\alpha\rho c_s^2+C
\end{equation}
where we have parametrized the viscous tensor as a function of the $\alpha$ 
parameter. From this expression it is possible to compute an accretion rate 
assuming that the viscous term vanishes at the inner edge of the disc, $R_0$ 
(which in general will be much shorter than our limit of resolution). After 
a $z$ integration, assuming that the disc is rotating with a Keplerian velocity 
and using the expression for $S_{r\theta}$
\begin{equation}
S_{r\theta}=\rho\nu\{[(\hat{r}\cdot\nabla)\vec{v}]\cdot\hat{\theta}+[(\hat{\theta}\cdot\nabla)\vec{v}]\cdot\hat{r}\}
\end{equation}
we obtain
\begin{equation}
\Sigma v_r v_\theta=\Sigma\nu\Omega+C.
\end{equation}

Denoting the disc angular velocity at the inner edge of the disc as 
$\Omega_0=(GM/R_0)^{1/2}$ with $M$ the mass of the central massive object
we have
\begin{equation}
\Sigma v_r R_0\Omega_0=C,
\end{equation}
and
\begin{eqnarray}
					 \Sigma v_r r-\Sigma v_r R_0^{1/2}r^{1/2}                         &=&\Sigma\nu\\
					 \Sigma v_r r\left[1-\left(\frac{R_0}{r}\right)^{1/2}\right]&=&\Sigma\nu\\
					 \dot{M}\left[1-\left(\frac{R_0}{r}\right)^{1/2}\right]     &=&2\pi\Sigma\nu.
\end{eqnarray}
As mentioned above, in our case $R_0<<r$ and so the mass accretion rate can be 
approximated by
\begin{equation}
					 \dot{M}\approx 2\pi\Sigma\nu.
\end{equation}
From the above equation and using the definition of the mass accretion rate it is 
possible to write $\nu=rv_r$ for this viscous MT model.

\begin{figure}
\centering
\includegraphics[width=1.0\columnwidth]{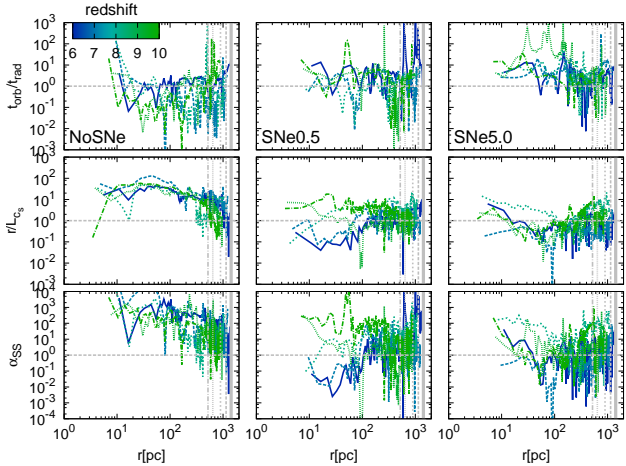}
\caption{From left to right: NoSNe, SNe0.5 and SNe5.0. In the top row, the ratio of the orbital to 
radial time as a function of radius for different redshifts. 
In the central row, the radius to pressure height scale ratio. 
In the bottom row, the $\alpha_{r\theta}$ parameter associated to the computed 
mass accretion rate in the simulations. From the top panel we can
deduce that the system has a non-stationary state with short radial times. 
The final $\alpha$ parameter can reach high values reflecting the non-stationary 
turbulent state of the system.}
\label{fig:alphaprof}
\end{figure}

Assuming that the disc is supported by the gas pressure in the vertical direction 
and writing the viscosity as $\nu=3\alpha L_{c_s}c_s/2$, with $L_{c_s}=r c_s/v_{\theta}$ 
we have:
\begin{eqnarray}
\dot{M} &=& \frac{3\pi r\Sigma\alpha c_s^2}{v_\theta} \nonumber \\
\dot{M} &=& \frac{3\pi \Sigma\alpha c_s^2}{\Omega}
\label{mdot}
\end{eqnarray}
Rearranging eq. \ref{mdot} it is possible to compute an $\alpha$ parameter as a 
function of the mass accretion rate in this viscous model as (see figure 
\ref{fig:alphaprof}):
\begin{equation}
\alpha = \frac{\dot{M}\Omega}{3\pi \Sigma c_s^2}
\end{equation}
Now, using the definition of the mass accretion rate we can write
\begin{eqnarray}
v_r   &=&\frac{3}{2}\frac{\alpha c_s^2}{v_\theta} \nonumber \\
\alpha&=&\frac{2}{3}\frac{v_r v_\theta}{c_s^2}\\
\alpha&=&\frac{2}{3}\frac{v_r}{v_\theta}\left(\frac{r}{L_{c_s}}\right)^2\nonumber \\
\alpha&=&\frac{1}{3\pi}\frac{t_{orb}}{t_{rad}}\left(\frac{r}{L_{c_s}}\right)^2
\end{eqnarray}

We emphasize that the previous expressions for $\dot{M}$ and $\alpha$ arise 
after assuming a stationary process for the momentum evolution, i.e. 
$\partial(\rho v_i)/\partial t=0$. This implies that the term inside the 
divergence (the rate of momentum fluxes) should be a constant. In our case, 
as can be seen in fig. \ref{fig:stressprof}, such an assumption is not valid 
due to the highly dynamic nature of the system: hierarchical mass assembly 
producing DM halo mergers, non-isotropic accretion due to the filamentary 
structure around the central DM halo and SNe explosions acting on the system.

\section*{Appendix 3}
\label{appendixD}
\begin{figure}
\centering
\includegraphics[width=1.0\columnwidth]{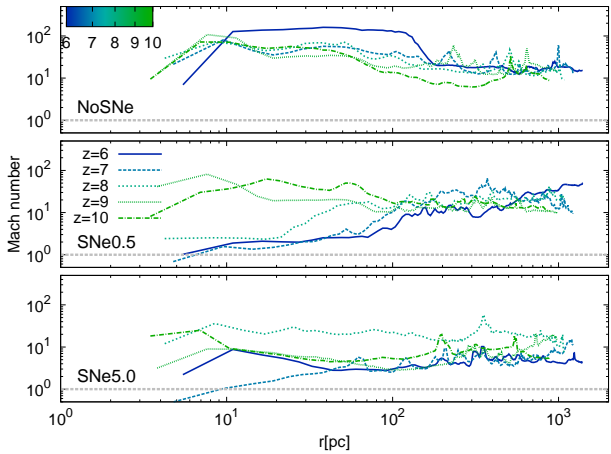}
\caption{Mach number as a function of radius for our simulations. The systems develop 
super-sonic velocities with high Mach numbers. This fact explains the high Reynolds 
stresses in all our galaxies.}
\label{fig:profmach}
\end{figure}

Figure \ref{fig:profmach} shows the Mach number of our three systems as a function of
radius for different redshifts. The Mach number is defined as $M=v_{rms}/c_s$.
Due to the violent conditions in high redshift galaxies our three experiments develop 
supersonic velocities. 

\section*{Appendix 4}
\label{appendixE}
\begin{figure}
\centering
\includegraphics[width=1.0\columnwidth]{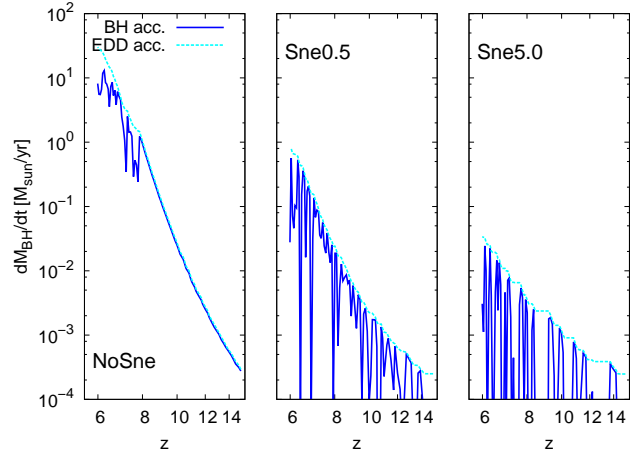}
\caption{BH mass accretion rat for our three simulations. From left to right: NoSNe, SNe0.5 and
SNe5.0. The dashed cyan line marks the Eddington accretion rate. The perturbation in our NoSNe run around $z=7$ is due to a number of mergers disturbing the galactic disc. SNe feedback
clearly affect the BH accretion reducing efficiently it growth.}
\label{fig:accrete}
\end{figure}

Figure \ref{fig:accrete} shows the BH mass accretion rate as a function of redshift for 
our three experiments. From this figure is clear the effect of SNe feedback on the
BH accretion rate.

\section*{Appendix 5}
\label{appendixF}
\begin{figure}
\centering
\includegraphics[width=1.0\columnwidth]{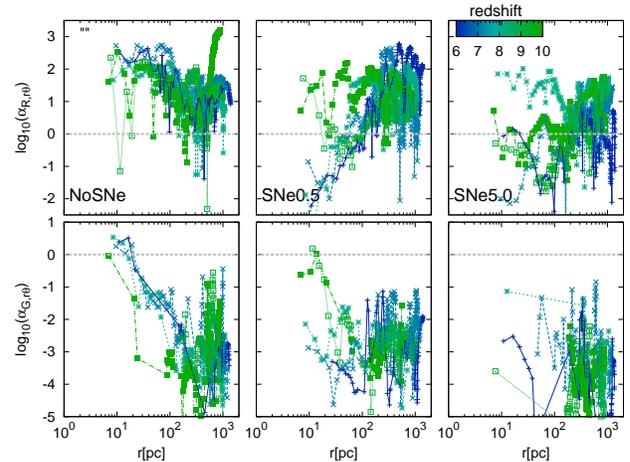}
\caption{Same as figure \ref{fig:stressprof} but for the alternative definition of 
$\alpha_{R,r\theta}$ from \citet{Hawley2000}.}
\label{fig:stressprof2}
\end{figure}

\citet{Hawley2000} proposes an alternative way to measure the Reynolds stress ``in 
terms of the difference between the total instantaneous angular momentum flux, and 
the mass flux times the average angular momentum'':
\begin{equation}
\langle R_{r\theta}\rangle=\langle \rho v_r v_\theta\rangle-\langle \rho v_r\rangle\langle \ell \rangle/r,
\end{equation}
with $\langle\ell\rangle$ the fluid specific angular momentum and $r$ the radial
coordinate. This quantity is shown in figure \ref{fig:stressprof2}. From this figure
it is clear that the Reynolds $\alpha$ has a similar behavior compared with 
eq. \ref{reynoldsstress}. Both expressions quantify the amount of angular momentum
flux due to perturbations in the azimuthal direction. The combination of large 
$\hat{\theta}$ perturbations and low sound speed (high Mach number systems) allows 
high Reynolds alpha parameters. 

\end{document}